\documentclass[11pt,a4 paper]{article}
\usepackage{float}
\usepackage{tocloft}
\usepackage{amsfonts,amsmath,amssymb}
\usepackage{multirow, verbatim}
\usepackage{shadow,enumerate}
\usepackage{makeidx}  % allows index generation
\usepackage{graphicx}
\usepackage{color}

\def\whitebox{{\hbox{\hskip 1pt
        \vrule height 6pt depth 1.5pt
        \lower 1.5pt\vbox to 7.5pt{\hrule width
                  3.2pt\vfill\hrule width 3.2pt}%
        \vrule height 6pt depth 1.5pt
        \hskip 1pt } }}
\def\qed{\ifhmode\allowbreak\else\nobreak\fi\hfill\quad\nobreak\whitebox\medbreak}
\date{}

\newcommand{\FF}{\mathbb{F}}

\newtheorem{example}{Example}

\newtheorem{proposition}{Proposition}
\newtheorem{remark}{Remark}

\newtheorem{op}{Open Problem}

\providecommand{\keywords}[1]{\textbf{\textit{Index terms---}} #1}

\begin{document}

\title{Optimizing the placement of tap positions and guess and determine cryptanalysis with variable sampling}

\author{ S. Hod\v zi\'c, E.~Pasalic, and Y. Wei%
\thanks{S. Hod\v zi\'c and E. Pasalic are with the University of Primorska, FAMNIT, Koper,
Slovenia (e-mails: samir.hodzic@famnit.upr.si, enes.pasalic6@gmail.com).}
%\thanks{Enes Pasalic is with the University of Primorska, FAMNIT, Koper,
%Slovenia (e-mail: enes.pasalic6@gmail.com)} }
\thanks{Y. Wei is with the Guilin University of Electronic Technology, Guilin, P.R. China (e-mail: walker$_{-}$wei@msn.com).}
}

\maketitle

\begin{abstract}
%\boldmath
\footnote{A part of this paper, related to algorithms for finding an optimal placement of tap positions, was presented at the First BalkanCryptSec conference, Istanbul, Turkey, October 2014.} In this article an optimal selection of tap positions for certain LFSR-based encryption schemes is investigated from both design and cryptanalytic perspective.  Two novel algorithms towards an optimal selection of tap positions are given which can be satisfactorily used to provide (sub)optimal resistance to some generic  cryptanalytic techniques applicable to these schemes. It is demonstrated that certain real-life ciphers (e.g. SOBER-t32, SFINKS and Grain-128), employing some standard criteria for tap selection such as the concept of full difference set,  are not fully optimized with respect to these attacks. These standard design criteria are quite insufficient and the proposed algorithms appear to be the only generic method for the purpose of (sub)optimal selection of tap positions. We also extend the framework of a generic cryptanalytic method called Generalized Filter State Guessing Attacks (GFSGA), introduced in \cite{EnesGFSGA} as a generalization of the FSGA method, by applying a variable sampling of the keystream bits  in order to retrieve as much information about the secret state bits as possible. Two different modes that use a variable sampling of keystream blocks are presented and it is shown that in many cases these modes may outperform the standard GFSGA mode.  We also demonstrate the possibility of employing GFSGA-like attacks to other design strategies such as NFSR-based ciphers (Grain family for instance) and filter generators outputting a single bit each time the cipher is clocked. In particular, when the latter scenario is considered, the idea of  combining GFSGA technique and algebraic attacks appears   to be a promising unified cryptanalytic method against NFSR-based stream ciphers.
%This possibility arises naturally due to the fact that GFSGA-like attacks reduce the preimage space of possible inputs to a filtering function using the knowledge of previous inputs, thus giving rise to the existence of low degree annihilators.
% achieved in combination with other cryptanalytic methods to other families of ciphers, thus not necessarily multiple output LFSR-based encryption schemes.

%\textcolor[rgb]{1.00,0.00,0.00}{Wei's comment: I believe that this work is very nice and solid. But I am feeling that it seems to be technological now!! (I guess)  if the abstract is interesting enough, then this article is half done. An important scientific problem (or issue) need to be originated. We need to convince the readers that we have solved this problem via novel idea, novel approaches, and new results. We need to try to look for a new expression  that is more concise, popular and easy to understand for general readers.  }

\end{abstract}
%\medskip
%\noindent


\section{Introduction}\label{intro}
Certain hardware oriented symmetric-key encryption schemes employ two basic primitives for generating keystream sequences, namely a linear feedback shift register (LFSR) and a nonlinear Boolean function. The LFSR is mainly used for storing the state bits and for providing the inputs to the Boolean function, which in turn processes these inputs  to output a single bit as a part of the keystream sequence. For a greater throughput,  a vectorial Boolean function, say $F$, may be used instead to provide several output bits at the time, thus $F:GF(2)^n\longrightarrow GF(2)^m$.
Nonlinear filter generator is a typical representative of a hardware oriented design in  stream ciphers. It consists of a single linear
feedback shift register (LFSR) and a nonlinear  function
$F:GF(2)^n\longrightarrow GF(2)^m$ that processes a fixed
subset of $n$ stages of the LFSR. This fixed subset of the
 LFSR's cells is usually called the {\em taps}.

The resistance of  nonlinear filter generators against various cryptanalytic attacks, such as (fast) correlation attacks \cite{Meier1,S85,Decim},  algebraic attacks \cite{CourMO,CourM03,MPCEC04}, probabilistic algebraic attacks \cite{AnBarkProb,PAAEP09}, attacks that take the advantage of the normality of Boolean functions \cite{Norm2012} etc., mainly depends on the choice of the filtering function $F$. The design rules  for ensuring good security margins against these attacks are more or less known today. Nevertheless, guess and determine cryptanalysis is a powerful cryptanalytic tool against these schemes whose efficiency is irrelevant of the cryptographic properties of the  filtering function (the same applies  to time-memory-data trade-off attacks  \cite{BirySha}, \cite{HellmTMTO}, \cite{HongS05}) but rather to the selection of LFSR: its size, primitive polynomial used and tapping sequence (tap positions used to supply $F$ with the inputs). The main goal of the guess and determine cryptanalysis, when applied to these schemes, is to recover  the secret state bits contained in the LFSR by guessing a certain portion of these bits and exploiting the structure of the cipher. The term structure here mainly refers to  the tap positions of LFSR used to provide the inputs to $F$  and  to some fixed positions of LFSR for implementing a linear recursion through the primitive connection polynomial. For the first time, it was explicitly stated in \cite{Gol96} that the choice of tap sequence may play more significant role than the optimization of $F$ in the context of inversion attacks introduced in \cite{Gol96}, see also \cite{Gol94,Gol2000}. To protect the cipher from  inversion attacks a full positive difference set was employed in \cite{Gol96}, where a set of positive integers $\Gamma=\{i_1,\ldots,i_n\}$ is called a full positive difference set if all
the positive pairwise differences between its elements are distinct. These sets are, for instance,
used in the design of self-orthogonal convolutional codes.

The basic idea behind  the attacks similar to inversion attacks is to exploit the shift of the secret state bits that are used as the input to the filtering function. To understand this assume that we sample the keystream bits at suitable time instances so that a portion of the  secret bits that are used in the previous sampling instance appear again as (part of) the input (at different positions) to the filtering function. Provided the knowledge of the output, this information then significantly reduces the uncertainty about the remaining unknown inputs. The designers, well aware of the fact that a proper tap selection  plays an important role in the design,  mainly use some standard (heuristic) design rationales such as taking the differences between the positions to be prime numbers (if possible), the taps are distributed over the whole LFSR etc.. Intuitively,  selecting  the taps
  at some consecutive positions of the LFSR should be avoided (see also \cite{Anders}), and similarly placing these taps at the positions used for the realization of the feedback connection
polynomial is not a good idea either.

 Even though a full positive difference set is a useful design criterion which ensures that there are no repetitions of several input bits, it is  quite insufficient  criterion which does not prevent from the attacks such as GFSGA (Generalized Filter State Guessing Attack)  introduced in \cite{EnesGFSGA}. For instance, assume for simplicity that the inputs to the filtering function are taken at tap positions ${\cal I}=\{ 3,6, 12, 24\}$ of the employed LFSR, thus our filtering function takes four inputs, i.e., $n=4$. It is easily verified that all the differences are distinct and the set of (all possible) differences is
$D^{{\cal I}}=\{i_j - i_k : i_j, i_k \in {\cal I}, i_j > i_k \} = \{3, 6, 9, 12, 18, 21 \}$. Nevertheless, all these numbers being multiple of 3 would enable an efficient  application of  GFSGA-like cryptanalysis  since the information about the previous states would be maximized.

Another  criterion considered in the literature, aims at ensuring that a multiset of differences of the tap positions is mutually coprime. This means, that for a given set of tap positions $\mathcal{I}=\{i_1,i_2,\ldots,i_n\}$ of an LFSR of length $L$ (thus $1 \leq i_1 < i_2 < \ldots <i_n \leq L$) all the elements in the difference set  $\mathcal{D}^{{\cal I}}=\{i_j-i_k: i_j, i_l \in \mathcal{I}, i_j >i_k\}$ are mutually coprime.
This condition, which would imply an optimal resistance to GFSGA-like methods,  is easily verified   to be impossible  to satisfy (taking any two odd numbers their difference being even would prevent from taking even numbers etc.). Therefore, only the condition that the consecutive distances are coprime appears to be reasonable, that is, the elements of $D=\{ i_{j+1} -i_j : i_j \in {\cal I}\}$ are mutually coprime. An exhaustive search is clearly infeasible, since  in real-life applications to select (say) $n=20$ tap positions for a driving LFSR of length 256  would give $\binom{256}{20}=2^{98}$ possibilities to test for optimality. %which is clearly infeasible.

This important issue of finding (sub)optimal solutions for selecting  tap positions, given the input size $n$ and the length $L$ of the driving LFSR, appears to be highly neglected in the literature. Certain criteria for the choice of tap positions was firstly mentioned in \cite{Gol96} but a more comprehensive treatment of this issue was firstly addressed recently in \cite{SHT}, where two algorithms for the purpose of selecting taps (sub)optimally were presented. These algorithms were used in \cite{SHT} to   show that the selection  of tap positions in real-life stream ciphers such as SOBER-t32 \cite{Hawkes} and SFINKS \cite{Sfinx} could have been (slightly) improved to ensure a better resistance of these ciphers to GFSGA-like cryptanalysis. For self-completeness and since this manuscript extends the work in \cite{SHT}, the above mentioned algorithms and the theoretical discussion for their derivation is also given, though in a slightly suppressed form.  The emphasis is given to the construction of algorithms and their relation to the criteria for tap selection proposed in \cite{Gol96}. It is shown that these criteria are embedded in our algorithms but they are not sufficient for protecting the considered encryption schemes against GFSGA-like methods adequately. In particular, the selection of tap positions for Grain-128 cipher is far from being optimized allowing for a significant improvement of its resistance to GFSGA-like attacks as shown in Section \ref{sec:applic}. Thus, these algorithms appear to be the only known efficient and generic method for the purpose of selecting tap positions (sub)optimally.

Another goal of this manuscript is to further extend the GFSGA framework  by considering a variable mode of sampling which was not addressed in FSGA \cite{EnFSGA} or GFSGA \cite{EnesGFSGA}. For a better understanding of the differences between various modes we give a brief description of the known modes, thus FSGA and GFSGA,  and discuss our extended mode of GFSGA which we describe later in more detail.
 The basic idea behind  FSGA is to recover secret state bits by reducing the  preimage space of the filtering function $F:GF(2)^n\longrightarrow GF(2)^m$ using the knowledge of previously guessed bits. Since for uniformly distributed $F$ there are $2^{n-m}$  preimages for any observed $m$-bit output block, the attacker may  for each  choice of $2^{n-m}$ many possible inputs (over the whole set of sampling instances) set up an overdefined  system of linear equations in
secret state bits. This attack turns out to be  successful only for relatively large $m$, more precisely for approximately $m > n/2$. In certain cases, the running time of  FSGA may be lower than the running time of a classical algebraic attack (cf. \cite{EnFSGA}). Nevertheless,  the placement of tap positions of a
nonlinear filter generator were of no importance for this attack. More precisely, only one bit of the information was considered to be known from the previous states in the case of FSGA. The complexity of the attack was significantly improved in \cite{EnesGFSGA}, where the information from the neighbouring taps, in the attack named GFSGA (Generalized FSGA), was used for a further reduction of the preimage space. In particular, the attack complexity of GFSGA is very sensitive to the  placement of taps, which essentially motivated the initiative taken in  \cite{SHT} for devising the algorithms for computing (sub)optimal choices of tap positions that give the maximum resistance against GFSGA.

However, even this generalized approach, which takes into account the tap positions of the driving LFSR,  turns out not to be fully optimal. The main reason is that GFSGA  works with a constant sampling rate and its optimal sampling rate can be  easily computed given the set of tap positions. The algorithms themselves, as presented in \cite{SHT}, are  designed to select tap positions that gives the largest resistance to GFSGA with a constant sampling rate. Therefore, a natural question that arises here regards the impact on the robustness  of the cipher if a variable sampling rate is used instead. This issue is elaborated here by introducing two different modes  of  GFSGA  with non-constant (variable) sampling rate. These modes are much less dependent on the choice of tap positions and in many cases their performance is demonstrated through examples to be better than that of the standard GFSGA.

We notice that the main difficulty, when comparing the performance of these modes theoretically, lies in the fact that there are intrinsic trade-offs between the main parameters involved in the complexity computation, cf. Remark \ref{tradeoff}. The main reason is that each of these modes attempt to reduce the preimage space based on the knowledge of some secret state bits that reappear as the inputs, but at the same time these linear equations (describing the known/guessed  secret state bits) have already been used for setting up a system of linear equations to be solved once the system becomes overdefined. Thus, increasing the number of repeated bits makes a reduction of the preimage space more significant (less bits needs to be guessed) but at the same time more sampling is required since the repeated bits do not increase the rank of the system of linear equations. This is the trade-off that makes the complexity analysis hard and consequently no theoretical results regarding the performance of the attack modes can be given.

Finally, well aware of the main limitation of GFSGA-like attacks, which are efficiently applicable to LFSR-based ciphers with filtering function $F:GF(2)^n \rightarrow GF(2)^m$ where $m >1$, we briefly discuss their application to single output filtering functions (thus   $m=1$) and to ciphers employing nonlinear feedback shift registers (NFSRs). In both cases we indicate that GFSGA attacks may be adjusted   to work satisfactory in these scenarios as well. Most notably, there might be a great potential in applying GFSGA attacks in combination with other cryptanalytic techniques such as algebraic attacks.  This possibility arises naturally due to the fact that GFSGA-like attacks reduce the preimage space of possible inputs to a filtering function using the knowledge of previous inputs, thus giving rise to the existence of low degree annihilators defined on a restriction of the filtering function (obtained by keeping fixed a subset of known input variables). In another direction, when considering NFSR-based ciphers  we propose  a novel approach  of mounting internal state recovery attacks on these schemes which employs the GFSGA sampling procedure but without solving the deduced systems of equations at all.  More precisely, this new type of internal state recovery attack  collects the outputs within a certain sampling window  which then enables an efficient recovery of  a certain portion of internal state bits. This is done  by filtering out the wrong candidates based on the knowledge of reduced preimage spaces that correspond  to the observed outputs.

The rest of the article is organized as follows. In Section~\ref{sec:prel}, some basic definitions regarding Boolean functions and the mathematical formalism behind the structure of nonlinear filter generators is given. For completeness, a  brief overview of FSGA and GFSGA is also given in this section.   Two different modes of  GFSGA with variable sampling distance are introduced in Section~\ref{sec:varstep}. The complexity analysis of these modes in terms of the number of repeated state bit equations is addressed here as well.  In Section~\ref{sec:comp}, the performance  of different attack modes and the algorithms for determining a (sub)optimal selection of tap positions are presented. The possibility of applying GFSGA to single output filtering functions and to NFSR-based ciphers is discussed in Section~\ref{sec:applic}. Some concluding remarks are given in Section~\ref{sec:conc}.

\section{Preliminaries}\label{sec:prel}

A Boolean function is a mapping from $GF(2)^n$ to $GF(2)$, where
$GF(2)$ denotes the binary Galois field and  $GF(2)^n$ is an $n$-dimensional vector space spanned over $GF(2)$. A  function $f:GF(2)^n \rightarrow GF(2)$   is commonly represented using its associated algebraic normal form (ANF) as follows:
\begin{eqnarray*}\label{f}
f(x_1,\ldots,x_n)={\sum_{u\in GF(2)^n}{\lambda_u}}{(\prod_{i=1}^n{x_i}^{u_i})},
\end{eqnarray*}
where $x_i\in GF(2)$, $(i=1, \ldots, n)$, ${\lambda_u \in GF(2)}$, $u=(u_1, \ldots,u_n)\in GF(2)^n$. A vectorial (multiple output) Boolean function $F(x)$ is a mapping from $GF(2)^n$ to $GF(2)^m$, with $m \geq 1$, which can also be regarded as a collection of $m$ Boolean functions, i.e., $F(x)=(f_1(x), \ldots, f_m(x))$. Commonly, $F(x)$ is chosen to be uniformly distributed, that is, $\# \{{x\in GF(2)^n \mid F(x)=z}\}=2^{n-m}$, for all $z\in GF(2)^m$. In some cases, we also use the notation $|A|$ to denote the cardinality of the set $A$. Moreover, for any $z=(z_1, \ldots, z_m)\in GF(2)^m$, we denote the set of preimage values by $S_z=\{x\in GF(2)^n \mid F(x)=z\}$.

\subsection{Nonlinear filter generator}%\label{sec:fsga}

A nonlinear filter generator \cite{Handb} consists of a single LFSR of length $L$ (thus comprising $L$ memory cells)
whose $n$ fixed positions (taps)   are used as the inputs to a filtering
 function $F:GF(2)^n \rightarrow GF(2)^m$.   These   $m$ outputs are used as a part of the keystream bits each time  the LFSR is clocked and the internal
state is then updated by a transition function. More formally,
$$(z_1^t, \ldots, z_m^t)=(f_1(\ell_n({\bf s}^t)), \ldots ,
f_m(\ell_n({\bf s}^t))),$$ where ${\bf s}^t=(s_0^t, \ldots,
s_{L-1}^t)$ is the secret  state of the LFSR at time $t$. The notation
$\ell_n({\bf s}^t)$ means that only a fixed subset of $n$ bits  of ${\bf s}^t=(s_0^t, \ldots,
s_{L-1}^t)$  is used as the input to Boolean functions $f_1, \ldots,f_m$, and $z_1^t, \ldots, z_m^t$ are the corresponding output keystream
bits. These fixed $n$ positions of LFSR, used as the input to $F$, are called {\em tap positions} in the sequel and will be denoted by ${\cal I}_0=\{l_1,l_2,\ldots,l_n\}$.

The LFSR is updated by  computing a newly generated (rightmost) bit $s_L^{t+1}$   as a linear combination of $s_0^t,\ldots, s_{L-1}^t$ determined by the connection polynomial and then shifting its content to the left. That is, its state is updated as $(s_1^{t+1},\ldots,s_{L-1}^{t+1},s_L^{t+1})\hookleftarrow (s_0^{t}, s_1^t\ldots,s_{L-1}^{t})$. Due to linearity of its feedback connection polynomial, at any $t \geq 0$ we have  $\ell_n(s_0^t, \ldots,
s_{L-1}^t)=(\psi_1^t({\bf s}^0), \ldots, \psi_n^t({\bf s}^0))$,
where the linear functions $\psi_i^t({\bf s}^0)=\sum_{j=0}^{L-1}{a_{i,j}^{t}s_j^0},$ $(i=1, \ldots, n)$, are unique linear combinations of the initial secret state bits ${\bf{s}}^0=(s_0^0,\ldots,s_{L-1}^0)$, at time $t=0$. The binary coefficients $a_{i,j}^{t}$ above can therefore be efficiently computed  from the connection polynomial of LFSR for all $t \geq 0$.

\subsection{An overview of FSGA and GFSGA}\label{sec:over}

For self-completeness and due to the close relation with subsequent sections, we briefly describe the main ideas behind FSGA and its extension GFSGA.
For both attacks there is no restriction on $F:GF(2)^n \rightarrow GF(2)^m$, thus $F$ satisfies all the relevant cryptographic criteria including  a uniform distribution of its preimages. This also indicates a generic nature of GFSGA and the possibility of improving its performance in case  the filtering function is not optimally chosen.
%\textbf{We need one such sentence, since at several places the 3rd reviewer is trying to comment some attacks which are bringing properties "on $F$"......we do no discuss $F$ at all in our article, i.e., there is no need to discuss "statistical distinguishers, conditional correlation attacks etc etc"}

\subsection{FSGA description}

For any observed keystream block $z^t=(z_1^t, \ldots,
z_m^t)$ at time $t$, there are $2^{n-m}$ possible inputs $x^t\in
S_{z^t}$. Moreover, for every guessed preimage $x^t=(x_1^t, \ldots,x_n^t)\in S_{z^t}$,
one obtains $n$ linear equations in the initial secret state bits $s_0^0,
\ldots, s_{L-1}^0$ through $x_i^t=\sum_{j=0}^{L-1}a_{i,j}^ts_j^0$, for
$1\leq i \leq n$.  The goal of the attacker is to recover the initial
state bits $(s_0^0, \ldots, s_{L-1}^0)$ after obtaining sufficiently many
keystream blocks $z^t=(z_1^t, \ldots, z_m^t)$. If the attacker observes the
outputs at some time instances $t_1, \ldots, t_c$, so that $nc>L$, then with high
probability  each system of $nc$ linear equations will be solvable but only one system will provide a  unique and consistent (correct) solution.

There are $2^{(n-m)c}$ possibilities of  choosing $c$ input tuples
$(x_1^{t_1}, \ldots,x_n^{t_1}), \ldots, (x_1^{t_c}, \ldots,x_n^{t_c})$ from $S_{z^{t_i}}$, and
 for each  such  $c$-tuple a system of $nc$ linear equations in $L$
variables (secret state bits) is obtained.
%  Especially, if the coefficient matrix
% for this system is full rank, then there is a uniqueness of the
% solution.
 The complexity of solving a single  overdefined system of
linear equations with $L$ variables is about $L^3$ operations.
Thus, the complexity of the FSGA is about
$2^{(n-m)c}L^3$ operations, where $c \approx \lceil \frac{L}{n}\rceil$.

\subsection{GFSGA description} \label{sec:gfsga}

The major difference to FSGA is that the GFSGA method efficiently  utilizes the tap positions of the underlying LFSR. Let the tap positions of the LFSR be specified by the set  $\mathcal{I}_0=\{l_1,l_2, \ldots, l_n \}$, $1\leq l_1 < l_2 < \ldots < l_n \leq L$. If at any time instance $t_1$, we assume that the content of the LFSR at these tap positions  is given by ${\bf s}_{\mathcal{I}_0}^{t_1}=(s^{t_1}_{l_1},\ldots,s^{t_1}_{l_n})$, then at $t=t_1+\sigma$ we have ${\bf s}_{\mathcal{I}_0+\sigma}^{t_1+\sigma}=(s^{t_1+\sigma}_{l_1+\sigma},\ldots,s^{t_1+\sigma}_{l_n+\sigma})$, where the notation ${\bf s}_{\mathcal{I}_0}^t$ means that we only consider the state bits at tap positions.
% where cutting modulo $L$ can be performed if necessary.
%  is shifted to the position $(i_1+i, i_2+i, \ldots, i_n+i)$ at the time instance $t+i,$ $i \geq 1$.  That is,  the  following description of the cipher is valid,
%\begin{eqnarray*}
%t &:& (a_1, a_2, \ldots, a_n)\rightarrow(i_1, i_2, \ldots, i_n),\\
%t+1 &:& (a_1, a_2, \ldots, a_n)\rightarrow(i_1+1, i_2+1, \ldots,
%i_n+1),\\
%%t+2 &:& (a_1, a_2, \ldots, a_n)\rightarrow(i_1+2, i_2+2, \ldots,
%%i_n+2),\\
%& & \vdots \\
%t+i &:& (a_1, a_2, \ldots, a_n)\rightarrow(i_1+i, i_2+i, \ldots,
%i_n+i).
%\end{eqnarray*}
Notice that the state bits at tap positions at time instance $t_1+\sigma$, denoted as $s^{t_1+\sigma}_{\mathcal{I}_0+\sigma}$, does not necessarily intersect with $s^{t_1}_{\mathcal{I}_0}=s^{t_1}$, thus if the intersection is an empty set no information from the previous sampling can be used at  $t_1 + \sigma$. The extreme case of a poor cryptographic design corresponds to the selection
$\mathcal{I}_0=\{l_1,l_2, \ldots, l_n \}=\{l_1,l_1+\sigma, \ldots, l_{n-1} + \sigma \}$, thus having $l_{i+1}=l_i + \sigma$ for any $i=1, \ldots,n-1$. This would imply that at $t=t_1 + \sigma$, based on the knowledge of the state bits ${\bf s}_{\mathcal{I}_0}^{t_1}$, we would dispose with $n-1$ input bits to $F$ and the unknown $n$-th input is easily determined upon the observed output.

Nevertheless, we can always select $\sigma$ so that at least one bit of information is conveyed.  More formally, if we denote the outputs by $z^{t_1},\ldots,z^{t_c}$ at  $t_1, \ldots, t_c$, where $t_i=t_1+(i-1)\sigma$ and $ 1\leq\sigma\leq (l_n-l_1)$, the sampling process at these time instances may give  rise to identical linear equations since the equations $x_i^{t_u}=\sum_{j=0}^{L-1}a_{i,j}^{t_u}s_j$ (where $1 \leq i \leq n$) may be shifted to $x_l^{t_{v}}=\sum_{j=0}^{L-1}a_{i,j}^{t_{v}}s_j$, for some $1\leq i<l\leq n, 1\leq u<v\leq c$. For simplicity, throughout the article the LFSR state bits at time instance $t_i$ are denoted by $ s^i=(s_{0+i},\ldots,s_{L-1+i})$, whereas for the state bits at tap positions we use the notation $ s^{t_i}=(s^{t_i}_{l_1},\ldots,s^{t_i}_{l_n}).$
%\begin{remark}\label{GFSGArem}
%Since the consecutive observed outputs $y^{t_{i-1}}$ and $y^{t_i}$, at time instances $t_{i-1}$ and $t_i$ respectively, differ for a step $\sigma,$ i.e. $t_i=t_{i-1}+\sigma,$ $i=1,2,\ldots,c,$ this mode of the GFSGA attack we call the GFSGA with a constant step, or shortly $GFSGA$. \textcolor[rgb]{0.98,0.00,0.00}{In Section \ref{sec:varstep} we introduce more general description of the GFSGA attack, where the sampling step $\sigma$ may actually take different values at every time instances $t_i,$ $i=1,2,\ldots,c.$ In that sense, we would like to distinguish GFSGA attacks with constant and non-constant (variable) steps of sampling.}
%\end{remark}

This mode of the GFSGA attack will be called the GFSGA with a constant sampling step, or shortly $GFSGA.$
It is of importance to determine  how many identical linear equations will be obtained for all the sampling instances $t_1, \ldots, t_c$. By introducing  $k=\lfloor \frac{l_n-l_1}{\sigma}\rfloor$, and for $\mathcal{I}_0=\{l_1,l_2,\ldots, l_n\}$ defined recursively:
\begin{eqnarray}\label{i}
 \mathcal{I}_1&\hskip -3mm=\hskip -3mm &\mathcal{I}_0 \cap \{l_1+\sigma, l_2+\sigma, \ldots,
l_n+\sigma\},\nonumber\\
\mathcal{I}_2&\hskip -3mm=\hskip -3mm & \mathcal{I}_1 \cup \{ \mathcal{I}_0 \cap \{l_1+2\sigma,
l_2+2\sigma, \ldots, l_n+2\sigma \}\},\nonumber\\
& & \vdots \\
\mathcal{I}_k&\hskip -3mm=\hskip -3mm & \mathcal{I}_{k-1} \cup \{
\mathcal{I}_0  \cap \{l_1+k\sigma, l_2+k\sigma, \ldots, l_n+k\sigma  \}\},\nonumber
\end{eqnarray}
the analysis in \cite{EnesGFSGA} showed that the complexity of the $GFSGA$ is closely related to the parameter  $r_i= \#\mathcal{I}_i$, where $i=1, \ldots, k.$
\begin{remark}
The above notation means that if for instance some $i \in \mathcal{I}_1 $ and therefore  $i \in \mathcal{I}_0 $, then  the state bit $s_i^{t_2}$ was used in the previous sampling since it was at  position $i -\sigma \in \mathcal{I}_0 $ at time $t_1$, where $t_2=t_1 + \sigma$. The idea is easily generalized for $\#\mathcal{I}_i=r_i$,  where $i=2, \ldots,k$.
\end{remark}
The number of identical equations obtained in \cite{EnesGFSGA} is given as follows. If $c\leq k$, then in total $\sum_{i=1}^{c-1}r_i$ identical linear equations  are obtained, whereas for $c>k$ this number is $\sum_{i=1}^{k}r_i+(c-k-1)r_k$. Note that in this case  $r_k=r_{k+1}=\cdots=r_{c-1}$ due to the definition of $k$,  which simply guarantees that after $k$ sampling instances the maximum (and constant) number of repeated equations is attained. Notice that if $c\leq k$, then there are $$2^{n-m}\times 2^{n-m-r_1}\times \ldots\times 2^{n-m-r_{(c-1)}}$$ possibilities of  choosing $c$ input tuples $(x_1^{t_1}, \ldots, x_n^{t_1})$, $\ldots$ , $(x_1^{t_c}, \ldots,x_n^{t_{c}})$. For each such choice, a system of  $nc-\sum_{i=1}^{c-1}r_i>L$  linear equations in $L$ state variables can be obtained, where the number of samples  $c$ ensures that the systems of equations are overdefined and there will be a unique consistent solution to these systems.
 Consequently, the time complexity of the attack for $c \leq k$, corresponding to solving $2^{n-m}\times 2^{n-m-r_1}\times \ldots\times 2^{n-m-r_{(c-1)}}$ many linear systems, was estimated as,
\begin{eqnarray} \label{eq:3}
T_{Comp.}^{c\leq k}&\hskip -2.5mm=\hskip -2.5mm &2^{n-m}\times 2^{n-m-r_1}\times \ldots\times 2^{n-m-r_{(c-1)}}\times L^3,
\end{eqnarray}
and similarly,
if $c>k$,
%then there are $2^{(n-m)}\times 2^{(n-m-r_1)}\times \ldots\times 2^{(n-m-r_k)}\times 2^{(n-m-r_k)\times (c-k-1)}$ possibilities  of  choosing $c$ input tuples $(x_1^{t^\ast_1}, \ldots,x_n^{t^\ast_1}), \ldots, (x_1^{t^\ast_c}, \ldots,x_n^{t^\ast_{c}})$.
 the time complexity for $c > k$ was given by
\begin{eqnarray}\label{eq:4} %\nonumber
T_{Comp.}^{c> k}&\hskip -2.5mm=\hskip -2.5mm &2^{n-m}\times 2^{n-m-r_1}\times \ldots\times 2^{n-m-r_k}\times  2^{(n-m-r_k)\times (c-k-1)}\times L^3 .
\end{eqnarray}
%Similarly to the FSGA, the precomputation time complexity of the GFSGA is about $2^n$ operations. The memory complexity is about $n2^n$ bits.
Notice that there is an intrinsic trade-off between the time complexity and the condition that $nc-\sum_{i=1}^{c-1}r_i>L$ (regarding the uniqueness of solution) through the parameters $c$ and $r_i$. Indeed, the time complexity is minimized  if  $c$ is minimized and $r_i$ are maximized, but the condition $nc-\sum_{i=1}^{c-1}r_i>L$ requires on contrary that  $c$ is maximized and $r_i$ are minimized, see also Remark \ref{tradeoff}.
\begin{remark}\label{rem1}
If $n-m-r_i \leq 0,$ for some  $i \in \{1, \ldots, k \}$, then the knowledge of these $r_i$ bits allows the attacker to uniquely identify  the exact  preimage value from the set of $2^{n-m}$ possible preimages, i.e., we assume  $2^{(n-m-r_i)}=1$ when $n-m-r_i \leq 0$.
\end{remark}
A complexity  comparison of FSGA, GFSGA and CAA (Classical algebraic Attack) for certain choices of tap position was given in \cite{EnesGFSGA} and in certain cases the GFSGA mode of attack outperformed both other methods.
 % The tap positions and the sampling difference $\sigma$ are given below: \\
%
%(1) $\{3, 8, 13, 16, 21, 29, 32, 37, 44, 52, 67, 79, 92, 106, 111,$ $125, 155\}$, $\sigma=5, c=23$.\\
%
%(2) $\{2, 7, 17, 25, 27, 31, 48, 58, 61, 73, 82, 91, 103, 115, 123, $ $134,146,$ $ 156\}$, $\sigma=3, c=20$.
%\begin{table}[H]
%\footnotesize
%\centering
%\caption{Complexity comparison for different $(n, m)$ and $(K=80, L=160)$.}
%\vskip 2.5mm
%\begin{tabular}{ccc}
% \hline \noalign{\smallskip}
%  % after \\: \hline or \cline{col1-col2} \cline{col3-col4} \ldots
%$(n, m)$ &  (17, 6) &  (18, 7)\\ \noalign{\smallskip}
% \hline \noalign{\smallskip} %\hline
%$FSGA$ &  $ 2^{123} $  &  $ 2^{121} $    \\ \noalign{\smallskip}
% \hline
% \noalign{\smallskip}
%$CAA$ &  $2^{75}$ & $2^{75}$ \\ \noalign{\smallskip}
% \hline
%  \noalign{\smallskip}
%\textbf{\emph{GFSGA}} & $2^{53.97}$ & $2^{64.97}$    \\
% \hline
%\end{tabular}
%%\caption{Complexity comparison for different $(n, m)$ and $(K=80, L=160)$.}
%\label{tab1}
%\end{table}
%\textbf{Maybe to delete the table and taps above? }
\begin{remark}\label{optstep}
Note that when the $GFSGA$ method is applied, the attacker chooses the sampling distance $\sigma$ for which the minimal complexity $T_{Comp.}$ of the attack is achieved. This distance is called an \emph{optimal sampling distance}, and its calculation is done by checking the complexities of the attack for all $\sigma \in \{1,\ldots, L\}$.
\end{remark}

\section{GFSGA with a variable sampling step}\label{sec:varstep}

In this section, we describe the GFSGA method with a variable  sampling step $\sigma$, which we denote as $GFSGA^*$. The whole approach  is quite similar to $GFSGA$, the main difference is that we consider outputs at any sampling distances, i.e., the observed outputs $z^{t_{i-1}}$ and $z^{t_i}$ ($i=1,\ldots,c$) do not necessarily  differ by a fixed constant value and consequently the sampling distances $\sigma_1,\ldots,\sigma_c$ are not necessarily the same. It turns out that this approach may give a significant reduction in complexity compared to the standard version of the attack, see Section~\ref{sec:comp}.

We first adopt some notation to distinguish between the two modes.
%and then we show the same procedure of the standard GFSGA attack  applies to the $GFSGA^*$ attack as well. In the latter case,
 % Here values (steps) $\sigma_i$ can be taken to be equal or different.
% More discussion about their connection with the complexity of the attack in general will be provided in Section \ref{sec:modes}.
The number of observed outputs for which an overdefined system is obtained is denoted by $c^*,$ the corresponding number of repeated bits by $R^*$ and the attack complexity by $T^*_{Comp.}.$ The outputs taken at  time instances  $t_i$ are denoted by $w^{t_i}$, where we use the variable sampling steps  $\sigma_i$ so that $t_{i+1}=t_{i}+\sigma_i,$ for $i=1,2,\ldots,c^*-1.$ These values $\sigma_i$ (distances between the sampled  outputs), are referred to as  the variable steps (distances). Throughout this article, for easier identification of repeated bits over observed LFSR states, the state bits $s_0,s_1,\ldots$ are represented via their indices in $\mathbb{N},$ i.e., $s_i \rightarrow (i+1)\in \mathbb{N}$ ($i\geq 0$). In other words, the LFSR state bits $s^i=(s_{0+i},s_{1+i},\ldots,s_{L-1+i})$ are treated as a set of integers given by
\begin{eqnarray}\label{int}
(s_{0+i},s_{1+i},\ldots,s_{L-1+i})\leftrightarrow \{1+i,2+i,\ldots,L+i\}.
\end{eqnarray}
The purpose of this notation is to simplify the formal definition of LFSR state bits at tap positions introduced in the previous section. In addition, it allows us  to easier track these bits and to count  the number of repeated bits using the standard concepts of union or intersection between the sets.
%since the LFSR states we now consider as sets of integers and thus we can use unions or intersections between the states (sets).
 Henceforth, the LFSR states $s^i$ we symbolically write as $s^i=\{1+i,2+i,\ldots,L+i\}$ ($i\geq 0$), and this notation applies to state bits at tap positions $s^{t_i}$. For clarity, a few initial steps of the process of determining the preimage spaces is given in Appendix.

It is not difficult to see that the equalities (\ref{i}), used in $GFSGA$ to determine the parameters $r_i=\#\mathcal{I}_i$, are special case of the equalities given as:
\begin{eqnarray}\label{ii}
 \mathcal{I}^*_1&\hspace{-3mm}=\hspace{-3mm}&\mathcal{I}_0 \cap \{l_1+\sigma_1, l_2+\sigma_1, \ldots,
l_n+\sigma_1\}=s^{t_1}\cap s^{t_2},\nonumber\\
\mathcal{I}^*_2&\hskip -2.5mm=\hskip -2.5mm &\{s^{t_1}\cup s^{t_2}\}\cap \{l_1+(\sigma_1+\sigma_2), \ldots, l_n+(\sigma_1+\sigma_2) \}=\{s^{t_1}\cup s^{t_2}\}\cap s^{t_3},\nonumber \\
%&\hspace{-3mm}=\hspace{-3mm}&\{s^{t_1}\cup s^{t_2}\}\cap s^{t_3},\nonumber\\
&  \vdots & \nonumber \\
\mathcal{I}^*_j&\hspace{-3mm}=\hspace{-3mm}&\{s^{t_1}\cup \ldots\cup s^{t_{j}}\}\cap  \{l_1+\sum_{i=1}^{j}\sigma_i,\ldots, l_n+\sum_{i=1}^{j}\sigma_i  \}= \bigcup^j_{i=1}s^{t_{i}}\cap s^{t_{j+1}}, \nonumber \\
%&\hspace{-3mm}=\hspace{-3mm}& \bigcup^j_{i=1}s^{t_{i}}\cap s^{t_{j+1}},\\
&  \vdots &  \nonumber \\
\mathcal{I}^*_{c^*-1}%&\hspace{-3mm}=\hspace{-3mm}& \{s^{t_1}\cup \ldots\cup s^{t_{c^*-1}}\}\cap  \{l_1+\sum_{i=1}^{c^*-1}\sigma_i,\ldots, l_n+\sum_{i=1}^{c^*-1}\sigma_i  \}\nonumber \\
&\hspace{-3mm}=\hspace{-3mm}& \bigcup^{c^*-1}_{i=1}s^{t_{i}}\cap s^{t_{c^*}},
\end{eqnarray}
where $s^{t_1},\ldots,s^{t_{c^*}}$ represents the LFSR state bits at tap positions at time instances $t_1,\ldots,t_{c^*}.$

In general, the number of repeated equations which corresponds to the outputs $w^{t_1},\ldots,w^{t_{c^*}}$ at variable  distances $\sigma_i$ (i.e., $w^{t_{i+1}}=w^{t_i+\sigma_i}$), can be calculated as

\begin{eqnarray}\label{qkk}
q_j=\#\mathcal{I}^*_j=\#\left\{\bigcup^{j}_{i=1}s^{t_i}\cap \{s^{t_1}+\sum^{j}_{i=1}\sigma_i\}\right\},
\end{eqnarray}
%\max_{1\leq\sigma_k \leq L}
where all steps $\sigma_i$ are fixed for $i=1,\ldots,j$ and $j=1,\ldots,c^*-1.$ Note that the sets $\mathcal{I}^*_{j-1}$ ($j\geq 1$) correspond to outputs $w^{t_j}$ (where $\mathcal{I}^*_{0}=\mathcal{I}_{0}\stackrel{(\ref{int})}{=}s^{t_1}$). The sampling  instances are given as $t_j=t_1+\sum^{j-1}_{i=1}\sigma_i$, or $t_j=t_{j-1}+\sigma_{j-1},$ where $t_{j-1}=t_1+\sum^{j-2}_{i=1}\sigma_i$ is fixed. Similarly to  the GFSGA  with a constant sampling distance, the attack complexity  is estimated as
\begin{eqnarray}\label{cvs}
T^*_{Comp.}&\hskip -2.5mm =\hskip -2.5mm &2^{n-m}\times 2^{n-m-q_1}\times \ldots\times 2^{n-m-q_{c^*-1}}\times L^3.
\end{eqnarray}
%\begin{remark}\label{rem3}
Remark \ref{rem1} also applies here, thus if $n-m-q_j \leq 0$ for some  $j \in \{1, \ldots, c^*-1\}$, then the knowledge of these $q_j$ bits allows the attacker to uniquely identify  the exact  preimage value of the observed output, i.e., we have  $2^{(n-m-q_j)}=1$ when $n-m-q_j \leq 0$.
%\end{remark}
Notice that if the sampling steps $\sigma_i$ are equal, i.e., they have a constant value $\sigma=\sigma_i,$ for $i=1,2,\ldots,c^*-1,$ we get $q_i=r_i,$ $c^*=c,$ $R^*=R$ and $T^*_{Comp.}=T_{Comp.}.$
\begin{remark}\label{tradeoff}
It was already mentioned  that the analysis of complexity $T^*_{Comp.}$ appears to be very difficult, mainly due to the following reasons. For fixed $m,n$ and $L,$ it is clear that there exists a trade-off between the parameters $q_j$ ($j=1,\ldots,c^*-1$) and $c^*$. More precisely, for larger values $c^*$ we have that $2^{n-m-q_j}>1$ which implies the  increase  of $T^*_{Comp.}$, unless $n-m-q_j\leq 0.$ For this reason, the optimal step(s) of the GFSGA attack (whether we consider a constant or variable mode of the attack) is the one which minimizes $c^*$ satisfying at the same time the inequality $nc^*-R^*>L$. Furthermore, in the case of the constant GFSGA mode, the parameter $c$ for which $nc-R>L$ holds is not known prior to the completion of  the  sampling process. This also holds for the variable GFSGA mode, if we fix a sequence of sampling steps in advance. This, in combination with \cite[Remark 3]{SHT}, give more insight how complicated the relation  between parameters $m,n,L,q_j,c^*$ and $T^*_{Comp.}$ is.
\end{remark}

\subsection{The number of repeated equations for $GFSGA^{*}$}\label{GFSGA*R}

The relation between the number of repeated equations and complexity in the case of $GFSGA$ has been analyzed in \cite{SHT}, where an alternative method for calculating the number of repeated equations has been derived. Similarly, in this section we derive an alternative method for calculating the number of repeated equations for  $GFSGA^*$ (Proposition \ref{vs}).
%, whose application can be found in Section \ref{sec:modes}. First we start with recalling some definitions from \cite{SHT}.

For a given set of tap positions $\mathcal{I}_0=\{l_1,l_2,\ldots,l_n\}$, let us consider the set of differences between the consecutive tap positions, i.e.,
$$D=\{\hskip 0.1cm d_j\hskip 0.1cm|\hskip 0.15cm d_j=l_{j+1}-l_j,\hskip 0.2cm j=1,2,\ldots,n-1\}.$$
%To illustrate this approach, let us consider the set $\mathcal{I}_0=\{l_1,\ldots,l_5\}$ ($n=5$). We have $D=\{\hskip 0.1cm d_j\hskip 0.1cm|\hskip 0.15cm d_j=l_{j+1}-l_j,\hskip 0.2cm j=1,2,3,4\}$ and

 Based on this set   the so-called scheme of all possible differences was defined in \cite{SHT} as $D^{{\cal I}_0}=\{l_{j}-l_k : l_j, l_k \in {\cal I}_0, l_j >l_k\}$ and
 % \cite{SHT}, cf. Table \ref{tab:n=5}.
%\begin{table*}[t]
%\footnotesize
%\centering
%\caption{The scheme of all possible differences for the set $D=\{d_1,d_2,d_3,d_4\}$.}
%\begin{tabular}{cllll}
% \hline  \noalign{\smallskip}
%  Row$\backslash$Columns & Col. 1 & Col. 2 & Col. 3 & Col. 4 \\ \noalign{\smallskip} \hline  \noalign{\smallskip}
%  Row 1 & $d_1=l_2-l_1$ & $d_2=l_3-l_2$ & $d_3=l_4-l_3$ & $d_4=l_5-l_4$ \\ \hline  \noalign{\smallskip}
%  Row 2 &$d_1+d_2=l_3-l_1$ & $d_2+d_3=l_4-l_2$ & $d_3+d_4=l_5-l_3$ &  \\ \noalign{\smallskip} \hline  \noalign{\smallskip}
%  Row 3 &  $d_1+d_2+d_3=l_4-l_1$ & $d_2+d_3+d_4=l_5-l_2$ &  &  \\ \noalign{\smallskip} \hline  \noalign{\smallskip}
%  Row 4 &  $d_1+d_2+d_3+d_4=l_5-l_1$ &  &  &  \\  \noalign{\smallskip}
%  \hline
%\end{tabular}
%\label{tab:n=5}
%\end{table*}
%In Table \ref{tab:n=5}, Column $1$ specifies the repetition of some equations at the tap position $l_1,$ Column $2$ gives the repetition of equations on $l_2,$ etc. Similarly, Row $1$ takes into account the consecutive repetitions from $l_{j+1}$ to $l_j,$ Row $2$ regards the repetition from $l_{j+2}$ to $l_j,$ etc.
 used in \cite{SHT} to calculate the number of repeated equations for GFSGA.
 %using the scheme of differences was also described in \cite{SHT}. Here we just state the main result.
\begin{proposition}\cite{SHT}\label{4.1} Let $\mathcal{I}_{0}=\{l_1,l_2,\ldots,l_n\}$ be a set of tap positions, and let
$$D=\{l_{j+1}-l_j\hskip 0.1cm |\hskip 0.1cm j=1,2,\ldots,n-1\}=\{d_1,d_2,\ldots,d_{n-1}\}.$$
The number of repeated equations is calculated as
\begin{eqnarray}\label{R}
R=\sum_{i=1}^{n-1}(c-\frac{1}{\sigma}\sum_{k=i}^{m}d_k),
\end{eqnarray}
where $\sigma \mid \sum_{k=i}^{m}d_k$ for some $m\in \mathbb{N},$ $i\leq m\leq n-1$ and $\frac{1}{\sigma}\sum_{k=i}^{m}d_k\leq c-1$. Moreover, if $\frac{1}{\sigma}\sum_{k=i}^{m}d_k\geq c,$
for some $1\leq i\leq n-1$, then $(c-\frac{1}{\sigma}\sum_{k=i}^{m}d_k)=0.$ This means that the repetition of the same equations (bits) starts to appear after the LFSR state $s^{t_c}$.
\end{proposition}
In the case of  $GFSGA^*$, the scheme of differences can also be used to calculate the number of repeated equations $R^*$. However, in this case the calculation is slightly more complicated  compared to GFSGA, due to the fact that we have a variable step of sampling.

To illustrate the difference, let us consider the set of tap positions given by $\mathcal{I}_0=\{3,5,10,14,16\}$ ($L=20$ and $n,m$ not specified).  The corresponding set of  consecutive differences is given as $D=\{2,5,4,2\}.$  The scheme of all differences related to $D^{{\cal I}_0}$ is given as:
\begin{table}[H]
\footnotesize
\centering
\caption{The scheme of all differences for $D=\{2,5,4,2\}$.}
\vskip 2.5mm
\begin{tabular}{l|cccc}
  \hline
   & Col. 1 & Col. 2 & Col. 3 & Col. 4 \\ \hline
  $l_{j+1}-l_j$ & 2 & 5 & 4 & 2 \\ \hline
  $l_{j+2}-l_j$ & 7 & 9 & 6 &  \\ \hline
  $l_{j+3}-l_j$ & 11 & 11 &  &  \\ \hline
  $l_{j+4}-l_j$ & 13 &  &  &  \\
  \hline
\end{tabular}
%\caption{The scheme of all differences for $D=\{3,4,1,2\}$.}
\label{tabD}
\end{table}
In addition, let us assume that the first two  steps of sampling (distances between observed outputs) are given as: $\sigma_1=5,$ $\sigma_2=2$. To find the number of repeated bits (equations), we use the recursion of the  sets $\mathcal{I}^*_k$ given by relation (\ref{ii}). Even though $c^*$ is the number of outputs for which an overdefined system can be set up,  our purpose is to demonstrate the procedure of finding repeated bits for  $\sigma_1,\sigma_2$.
%sampling of repeated bits over the several observed outputs (i.e., we do not consider all sets $\mathcal{I}^*_0,\ldots,\mathcal{I}^*_{c^*-1}$ for some $c^*$, but only a few initial steps of sampling for taken $\sigma_1,\sigma_2,\sigma_3$).
 The computation of the number of repeated bits is  as follows:\\
1) The  state bits at tap positions at time $t_1$ correspond to $\mathcal{I}_0=\{l_1,l_2,l_3,l_4, l_5\}=\{3,5,10,14,16\}$, thus $s^{t_1}=(s_{2},s_4,s_{9},s_{13},s_{15})\stackrel{(\ref{int})}{=}\mathcal{I}_0$.
% it means that the preimage space $S_{w^{t_1}}$ has the size $2^{n-m}$ (\textbf{Step 1}). Recall that the numbers in the set $s^{t_1}=\mathcal{I}_0$, as well as in all its shifts $s^{t_1+\sigma_i}$ by variable steps $\sigma_i$, correspond to bits in LFSR states $s^{t_i}$.
 Since the first sampling distance $\sigma_1=5,$ we consider the LFSR state $s^{t_2}=\{\mathcal{I}_0+\sigma_1\}$ which is given as  $$s^{t_2}=\{\mathcal{I}_0+5\}=(s_7,s_{9},s_{14},s_{18},s_{20})\stackrel{(\ref{int})}{=}\{8,10,15,19,21\}.$$
We obtain that $\mathcal{I}^*_1=s^{t_1}\cap s^{t_2}=\mathcal{I}_0 \cap (\mathcal{I}_0+5)=\{10\},$ which means that $1=\# \mathcal{I}^*_1=q_1$ bit is repeated  and found in $s^{t_2}$ from the first state $s^{t_1}=\mathcal{I}_0$.
% to second observing state $s^{t_2}=\mathcal{I}_0+5$ (it is the bit which corresponds to $10$). Notice that the number (bit) $10$ on the state $s^{t_1}$ was on tap position $l_3,$ but on the state $s^{t_2}$ it is on the tap position $l_2.$ Thus, we have a repeated bit from $l_3$ to $l_2$ tap position.

 In terms of the scheme of differences, this repetition corresponds to $d_2=l_3-l_2=\sigma_1=5$, found  as the first entry in Col. 2.  In addition, note that $d_2=5$ is the only entry in the scheme of differences $D^{\mathcal{I}_0}$ which is equal to $\sigma_1.$
 % Consequently, the preimage space $S_{w^{t_2}}$ which corresponds to the second observed output $w^{t_2}=w^{t_1+\sigma_1}$ has the size $2^{n-m-q_1}=2^{n-m-1}.$
The main difference  compared to GFSGA is that in this case we do NOT consider the divisibility by $\sigma_1$  in the scheme of differences due to variable sampling steps.

2) For $\sigma_2=2$,  the observed outputs $w^{t_2}$ and $w^{t_3}$ satisfy $w^{t_3}=w^{t_2+\sigma_2}=w^{t_1+\sigma_1+\sigma_2}.$ The LFSR state $s^{t_3}$, which corresponds to the output $w^{t_3}$, is given as $s^{t_3}=\{s^{t_2}+\sigma_2\}=\{s^{t_1}+(\sigma_1+\sigma_2)\}$, and therefore $$s^{t_3}=\{s^{t_2}+2\}=(s_{9},s_{11},s_{16},s_{20},s_{22})=\{10,12,17,21,23\}.$$
At this position we  check whether there are repeated bits from the LFSR state $s^{t_2},$ but also from the state $s^{t_1}$. To find all bits which have been repeated from the state $s^{t_2},$ we consider the intersection $s^{t_3}\cap s^{t_2}=\{10, 21 \}.$ In addition, the bits which are repeated from the state $s^{t_1}$ are given by the intersection $s^{t_3}\cap s^{t_1}=\{10\}.$ Hence, we have the case that the same bit, indexed  by  $10,$ has been shifted from the state $s^{t_1}$ (since $\sigma_1 + \sigma_2=7$ so that 3 + 7=10) and from $s^{t_2}$ (since $\sigma_2=2$) to $s^{t_3}.$ On the other hand, the intersection corresponding to 21 gives us an equation that has not been used previously. Thus, the number of known state bits used in the reduction of the preimage space is 2, and therefore $|S_{w^{t_3}}|=2^{n-m-2}.$
%On the other hand, there are only two equations that may be used in setting up a linear system since one bit (equation) is simply repeated.
%More precisely, on state $s^{t_1}$ we have that the number $10$ is on tap position $l_3$ and on state $s^{t_2}$ it is on the tap position $l_2.$ Clearly, in this case the number of repeated bits in total is only one, since we do not write the same equations (bits) in our system of equalities (which we make out of the set of repeated bits in our attack). Thus, we have $|S_{w^{t_3}}|=2^{n-m-1}.$
In terms of the scheme of differences, one may notice that we have $d_1=l_2-l_1=2=\sigma_2$ (which refers to repetition from $s^{t_2}$ to $s^{t_3}$) and which gives $d_1=2$ in Col. 1 and Col. 4. On the other hand,  for $d_1+d_2=l_3-l_1=7=\sigma_1+\sigma_2$ (which refers to repetition from $s^{t_1}$ to $s^{t_3}$)  the repeated bit is found in the same column, namely Col. 1, and therefore it is not counted.
%Let us denote $\sigma^{(2)}=\sigma_1$ and $\sigma^{(1)}=\sigma_2+\sigma_1.$
In general, we conclude here that if we have a matching of $\sigma^{(2)}=d_2$ and $\sigma^{(1)}=d_1+d_2$ with some numbers (entries) in the scheme of differences which are in the same column (as we have here $d_1=\sigma^{(2)}$ and $d_1+d_2=\sigma^{(1)}$), we calculate only one repeated bit in total from this column. If there were more than 2 matchings, the same reasoning applies and thus we would calculate only one repeated bit.\\\\
%\end{enumerate}
The procedure above may continue for any number of observed outputs at any sampling distances. For instance, if we consider some sampling step $\sigma_j$ ($j\geq 1$), then we also need to consider all sums of the steps $\sigma^{(j-i)}=\sum_{h=j-i}^{j}\sigma_h,$ for all $i=0,\ldots,j-1$ ($j\geq 1$) and their matchings with some entries in the scheme of differences. In general, every repeated bit means that some sum(s) of steps $\sigma^{(j-i)}=\sum_{h=j-i}^{j}\sigma_h,$ $i=0,\ldots,j-1$ is equal to some difference $\sum_{p=r}^{m}d_p,$ for some $m\in\mathbb{N},$ over different columns, where $r=1,\ldots,n-1$ relate to the columns in the scheme of differences. A total number of repeated bits $R^*$ is the sum of all repeated bits over observed outputs at distances $\sigma_j$ ($j\geq 1$). Note that the method for calculation of the number of repeated bits described above actually generalizes Proposition \ref{4.1}, since  the use of constant sampling distance is just  a special case of  variable  sampling.
\begin{proposition}\label{vs} Let $\mathcal{I}_{0}=\{l_1,l_2,\ldots,l_n\}$ be a set of tap positions, and let
$$D=\{l_{i+1}-l_i\hskip 0.1cm |\hskip 0.1cm i=1,2,\ldots,n-1\}=\{d_1,d_2,\ldots,d_{n-1}\}.$$
Denoting $\sigma^{(j-i)}=\sum_{h=j-i}^{j}\sigma_h,$ $i=0,\ldots,j-1,$ the number of repeated equations is calculated as
\begin{eqnarray}\label{R^*}
R^*=\sum_{j=1}^{c^*-1}q_j=\sum_{j=1}^{c^*-1}\left(\sum_{i=0}^{j-1}\frac{1}{\sigma^{(j-i)}}\sum_{p=r}^{m}d_p\right),
\end{eqnarray}
where the term $\frac{1}{\sigma^{(j-i)}}\sum_{p=r}^{m}d_p=1$ if and only if  $\sigma^{(j-i)}=\sum_{p=r}^{m}d_p$ for some $m,r\in \mathbb{N}$, $1\leq r\leq m\leq n-1$, otherwise it equals  $0.$ If for a fixed $r$ and different numbers $m$ we have more matchings $\sigma^{(j-i)}=\sum_{p=r}^{m}d_p$, then only one bit will be taken in calculation.
\end{proposition}
Clearly, the numbers $m$ and $r$ depend on the values $\sigma^{(j-i)},$ $i=0,\ldots,j-1$ ($j\geq 1$) since we only consider those numbers $m, r \in \mathbb{N}$ such that $\sum_{p=r}^{m}d_p$ is equal to $\sigma^{(j-i)}.$

\subsection{Two specific modes of $GFSGA^*$}\label{sec:modes}

In this section we present two modes of $GFSGA^*,$ which in comparison to $GFSGA$  depend less on the choice of  tap positions. First we start with a general discussion regarding the attack complexity.

In order to obtain a minimal complexity of the $GFSGA^*$ attack, it turns out that the main problem is actually a selection of the cipher outputs. This problem is clearly equivalent to the problem of selecting the steps $\sigma_i$ which gives the minimal complexity $T^*_{Comp.}.$ The number of repeated bits (equations) at  time instance $t_i$ (for some $i>1$) always depends on all previous sampling points at  $t_1,\ldots,t_{i-1}.$ This property directly follows from (\ref{ii}), i.e., from the fact that we always check the repeated bits which come from the tap positions of LFSR at time instances $t_1,\ldots,t_{i-1}.$ This means that the number of repeated bits $q_i$ at time instances $t_i,$ given by (\ref{qkk}), always depends on the previously chosen steps $\sigma_1,\ldots,\sigma_{i-1}.$
%In general, this is the main reason why both Proposition \ref{4.1} and Proposition \ref{vs} still require $c$ and $c^*$ respectively, besides the fact that they exclude the parameter $k$ introduced in Section \ref{sec:gfsga}.
This also implies that we cannot immediately calculate the number of required keystream blocks $c^*$ for which  the inequality $nc^*-R^*>L$ is satisfied. This inequality  can only be verified subsequently, once  the sampling distances and the number of outputs $c^*$ have been specified.
%once we start the sampling of repeated bits over outputs $w^{t_1},\ldots,w^{t_{c^*}}$ (at variable distances), we are not able to predict or calculate the number of outputs $c^*$ with Proposition \ref{vs} (neither with (\ref{ii})), for which  the inequality $nc^*-R^*>L$ is satisfied.
Therefore we pose the following problem.
\begin{op}\label{op1}
For a given set of tap positions $\mathcal{I}_0=\{l_1,\ldots,l_n\}$, without the knowledge of $c*$, determine an  optimal sequence of sampling distances $\sigma_i$  for which the minimal complexity $T^*_{Comp.}$ is achieved.
\end{op}
In what follows we provide two modes of the $GFSGA^*$ whose performance will be discussed in Section \ref{sec:comp}.

\subsection{$GFSGA^{*}_{(1)}$ mode of  attack}

In order to minimize the complexity $T^*_{Comp.}$, one possibility is to maximize the values $q_j$, given by (\ref{qkk}), by choosing suitable $\sigma_i.$ However, this approach implies a trade-off between the values $q_j$ and $c^*$,  since $c^*$ is not necessarily minimized.
% we do not know how many cipher outputs $c^*$ will be sufficient to satisfy the inequality $nc^*-R^*>L$ (to provide an overdefined system).
 More precisely:\\
1) For the first step $1\leq \sigma_1\leq L$ we take a value for which $q_1$ is maximized, i.e., for which the cardinality
$$q_1=\#\{s^{t_1}\cap (s^{t_1}+\sigma_1)\}=\#\{s^{t_1}\cap s^{t_2}\}$$ is maximized. Without  loss of generality, we can take the minimal $\sigma_1$ for which this holds.\\
2) In the same way, in the second step we take a value $1\leq \sigma_2\leq L$  for which
$$q_2=\#\{(s^{t_1}\cup s^{t_2})\cap s^{t_3}\}=\#\{(s^{t_1}\cup s^{t_2})\cap ((s^{t_1}+\sigma_1)+\sigma_2)\}$$ is maximized. As we know, the step $\sigma_1$ here is fixed by the previous step. We continue this procedure until an overdefined system is obtained.\\\\
%\end{enumerate}
In other words, the values $q_j$ are determined by the $maximum$ function over $\sigma_i,$ for $1\leq \sigma_i \leq L,$ i.e., we choose the steps $\sigma_i$ for which we have:
\begin{eqnarray}\label{qkk2}
q_j=\max_{1\leq \sigma_j \leq L}\#\left\{\bigcup^{j}_{i=1}s^{t_i}\cap \{(s^{t_1}+\sum^{j-1}_{i=1}\sigma_i)+\sigma_j\}\right\},
\end{eqnarray}
where $\sum^{j-1}_{i=1}\sigma_i$ is fixed and $j=1,\ldots,c^*-1$. Hence, the function $\max_{1\leq \sigma_j \leq L}$ used in (\ref{qkk2}) means that we are choosing $\sigma_i$ so that the maximal intersection of $s^{t_{j+1}}$ with all previous LFSR states $s^{t_{1}},\ldots,s^{t_{j}}$ (in terms of cardinality)  is achieved. This mode of $GFSGA^*$ we denote by $GFSGA^{*}_{(1)}.$

\subsection{$GFSGA^{*}_{(2)}$ mode of  attack}

Another mode of  $GFSGA^*$, based on the use of sampling distances that correspond to the differences between consecutive tap positions,  is discussed in this section.
 % In what follows we provide a description
The  selection of steps $\sigma_i$ and the calculation of repeated bits is performed as follows..

For a given set of tap positions $\mathcal{I}_0=\{l_1,\ldots,l_n\},$ let $D=\{d_1,\ldots,d_{n-1}\}$ be the corresponding set of differences between the consecutive tap positions. The sequence of sampling distances $\sigma_i$ between the observed outputs $w^{t_{i}}$ and $w^{t_{i+1}}$ is defined as:
\begin{eqnarray}\label{si}
\left\{\begin{array}{c}
        \sigma_{1+p(n-1)}=d_1, \\
        \sigma_{2+p(n-1)}=d_2, \\
        \vdots\\
        \sigma_{n-1+p(n-1)}=d_{n-1},
       \end{array}
\right.
\end{eqnarray}
%\begin{eqnarray}\label{si2}
%\sigma_{j+p(n-1)}=d_j,\;\; for \;\; j=1,2,\ldots,n-1, \\
%\end{eqnarray}
%where $p=0,1,2,\ldots.$
for $p=0,1,2,\ldots.$ That is, the first  $n-1$ sampling distances are taking values exactly from the set $D$  so that  $\sigma_1=d_1, \sigma_2=d_2,\ldots, \sigma_{n-1}=d_{n-1}.$ Then, the next  $n-1$ sampling distances  are again $\sigma_{n}=d_1, \sigma_{n+1}=d_2,\ldots, \sigma_{2n-2}=d_{n-1},$ and so on. This mode of the $GFSGA^*$ we denote as $GFSGA^{*}_{(2)}.$ For this mode, using Proposition \ref{vs},  we are able to calculate  a lower bound on the number of repeated equations for every sampling step. Recall that at some sampling instance $t_j$ there are some  repeated bit(s) if and only if  $\sigma^{(j-i)}=\sum^{j}_{h=j-i}\sigma_h,$ $i=0,1,\ldots,j-1,$ is equal to some $\sum^{m}_{p=r}d_p=l_m-l_r,$ for some $1\leq r\leq m \leq n-1.$ In addition, if in the same column in the scheme of differences (which is equivalent to considering a fixed $r$ and all the values $m\geq r$) we have more matchings $\sigma^{(j-i)}=\sum^{m}_{p=r}d_p,$ then only one bit is counted.

Hence, taking the first $n-1$ sampling steps to be  $\sigma_1=d_1, \sigma_2=d_2,\ldots, \sigma_{n-1}=d_{n-1},$ the scheme of differences (constructed for our set $D=\{d_1,\ldots,d_{n-1}\}$) and Proposition \ref{vs} imply that $q_1\geq 1$, since at least $\sigma_1=d_1$ is in Col. 1. Then $q_2\geq 2$, since we have $\sigma^{(2-0)}=\sigma^{(2)}=\sigma_2$ is equal to $d_2$ in Col. 2., and $\sigma^{(2-1)}=\sigma^{(1)}=\sigma_1+\sigma_2$ is equal to $d_1+d_2$ in Col. 1.
 Continuing this process we obtain $q_3\geq3,\ldots, q_{n-1}\geq n-1$, which in total gives at least $$1+2+\ldots+(n-1)=\frac{n(n-1)}{2}$$ repeated equations for the first $n-1$ observed outputs. In the same way, at least $\frac{n(n-1)}{2}$ of bits are always repeated if further sampling  at  $n-1$ time instances is performed in accordance to (\ref{si}). For instance, when $p=1$ in (\ref{si}) we have $\sigma_{n}=d_1.$
% By the previously taken steps $\sigma_i=d_i,$ $i=1,\ldots,n-1,$ among the sums $\sigma^{(n-i)}$ for $i=0,1,\ldots,n-1,$ at least $\sigma^{(n)}=\sigma_n=d_1=l_2-l_1$ implies that $q_n\geq 1.$ Note that here $\sigma^{(2)}$, which is equal to $d_1+d_2+\ldots+d_{n-1}=l_n-l_1$, does not imply the second repeated bit on $l_1$, since it has already been counted from $l_2.$ Clearly, the bit counted from $l_2$ is in fact the bit which had been on the tap position $l_n$ in some of the previous LFSR states.
 In general, for $1\leq i\leq n-1$ and $p\geq 0$ we have $q_{i+p(n-1)}\geq i$, where the sampling steps are defined by (\ref{si}). We conclude this section with the following remarks.
\begin{remark}\label{Mrem1}
Both $GFSGA^{*}_{(1)}$ and $GFSGA^{*}_{(2)}$  depend less on the placement of the tap positions in comparison to $GFSGA.$ Indeed, for both modes the sampling distances $\sigma_i$ are selected with respect to a given placement of  tap positions but regardless of what this placement in general might be. These modes are therefore more useful for cryptanalytic purposes rather than to be used in the design of an optimal allocation of tap positions (given the length of LFSR and the number of taps).
\end{remark}
\begin{remark}\label{Mrem2}
In the case of $GFSGA$ where we have equal distances between the observed outputs, one may notice that the sequence of numbers $r_i=\#\mathcal{I}_i$ is always an increasing sequence. On the other hand, the sequence of numbers $q_i=\#\mathcal{I}^*_i$ for $GFSGA^*$ may not be increasing at all. In connection to Open problem \ref{op1}, neither $GFSGA^{*}_{(1)}$ nor $GFSGA^{*}_{(2)}$ automatically provides an optimal sequence of steps $\sigma_i$ (which would imply the minimization of $T^*_{Comp.}$). This means that there exist cases in which any of the modes $GFSGA$, $GFSGA^{*}_{(1)}$ and $GFSGA^{*}_{(2)}$ may outperform the other two (cf. Table \ref{tabCC} and Table \ref{tabCC2}, Section \ref{sec:cons}).
%For more details see \textcolor[rgb]{0.98,0.00,0.00}{Appendix.}
\end{remark}

\section{Comparision between $GFSGA$, $GFSGA^{*}_{(1)}$ and $GFSGA^{*}_{(2)}$}\label{sec:comp}

In this section we compare the performance of the  three GFSGA modes   when the tap positions are selected (sub)optimally by using the algorithms originally proposed in \cite{SHT}.  Moreover,  the case when the set of differences $D$  forms a full positive difference set is also considered and compared to the algorithmic approach.

\subsection{Overview of the algorithms for tap selection}\label{algs}

As briefly mentioned  in the introduction the concept of a full difference set, which ensures that all the entries in the set of all pairwise differences are different, is not a very useful criterion for tap selection. This is especially true when GFSGA-like cryptanalysis is considered as shown in Section \ref{sec:cons}. The same applies to the set of consecutive differences which may be taken to have mutually coprime entries which still does not ensure a sufficient cryptographic strength. Thus, there is a need for a more sophisticated algorithmic approach for designing (sub)optimally  the placement of $n$ tap positions for a given length $L$ of the LFSR. The main idea behind the algorithms originally proposed in \cite{SHT}, presented below for self-completeness, is  the use of the standard $GFSGA$ mode  with a constant sampling rate for the purpose of finding (sub)optimal placement of tap positions.

% The $GFSGA$ is used as a main tool to measure the quality of tap positions, although there are other important elements such as permutations and specific constrains of parameters $L,$ $n$ and $m.$ Now we recall a full description of the algorithms and discuss why the $GFSGA$, used as a measuring tool, should not be replaced by any of the previously introduced modes with variables steps.

The proposed algorithms  for the selection of tap positions use the  design rationales that maximize the resistance of the cipher to the standard mode of $GFSGA.$ Instead of specifying the set ${\mathcal I}_0$, both  methods aim at  constructing the set $D$ of consecutive differences which gives a low number of repeated equations (confirmed by computer simulations) for any  constant sampling distance  $\sigma$, which implies a good resistance to GFSGA-like methods. As an  illustration of the above mentioned algorithms  several examples were provided in \cite{SHT} and in particular an application to the stream cipher SOBER-t32 \cite{Hawkes} and SFINKS \cite{Sfinx} were analyzed in details. In what follows, we briefly recall the algorithms from \cite{SHT} and then we discuss their structural properties related to the presented modes of GFSGA as well as to the criteria for tap selection proposed in \cite{Gol96} which provides a resistance to inversion attacks.

%\textcolor[rgb]{0.98,0.00,0.00}{In order to construct a set of consecutive differences $D$, one may notice that algorithms proposed in \cite{SHT} require minor manipulation of input values (differences). Manipulation of input values itself is driven by the general rule which is denoted by \textbf{Step A}, that is, to select co-prime differences between the tap positions together with its placing over the whole cipher. Using this rule, two algorithms are discussed.}

In both algorithms a quality measure for the choice of tap positions  is the request that the optimal step (see Remark \ref{optstep}) of the $GFSGA$ attack  is as small as possible. The selection of tap positions itself is governed by the general rule, which is achieving co-prime differences between the tap positions (\textbf{Step A} in \cite{SHT}) together with distributing taps all over the register (thus maximizing $\sum^{n-1}_{i=1}d_i\leq L-1$, where $d_i=l_{i+1}-l_i$, $\mathcal{I}_0=\{l_1,\ldots,l_n\}$ being a set of tap positions).

The first algorithm is designed to deal with situations when the size $D$ is not large (say $\# D \leq 10$). In this case  the algorithm performs an exhaustive search of all permutations of the set $D$ (\textbf{Step B} in \cite{SHT}) and gives as an output a permutation which ensures a maximal resistance to $GFSGA.$ The complexity of this search is estimated as $O(n!\cdot K),$ where $K$ corresponds to the complexity of calculation  $T_{Comp.}$ for all possible $\sigma$.
\begin{remark}\label{rsigma}
As mentioned in \cite{SHT}, to measure the quality of a chosen set of differences $D$ with respect to the maximization of $T_{Comp.}$ over all  $\sigma$, the computer simulations indicate that  an optimal ordering of the set $D$ implies a small value of  an optimal sampling distance $\sigma$. When choosing an output permutation (cf. Step 5 below), we always consider both $\sigma$ and $T_{Comp.}$ though  $\sigma$ turns out to be a more stable indicator of the quality of a chosen set $D.$ %Contrary, it must not be the case.
\end{remark}
For practical values of $L$, usually taken to be $L=256$, the time complexity of the above algorithm becomes practically infeasible already for $n>10$. Using the previous algorithm, the second algorithm proposed in \cite{SHT} regards a case when $\# D$ is large. To reduce its factorial time complexity, we use the above algorithm only to process separately the  subsets of the multiset $D$  within the feasibility constraints imposed on the cardinalities of these subsets. The steps of this modified algorithm are given as follows.
\bigskip
\noindent
%{\bf INPUT:} The numbers $n,m,L$.\\
%{\bf OUTPUT:} The set $D$ with the best arrangement of its elements.

{\bf STEP 1:} Choose a set $X$ by \textbf{Step A}, where $\#X<\#D$ for which \textbf{Step B} is feasible;

{\bf STEP 2:} Find the best ordering of $X$ using the algorithm in \textbf{ Step B} for

{\hspace{1.5cm} $L_{X}=1+\sum_{x_i\in X}x_i<L$  and $m_{X}=\lfloor\#X\cdot\frac{m}{n-1}\rfloor$;

{\bf STEP 3:} Choose a set $Y$ by \textbf{Step A}, where $\#Y<\#D$ for which \textbf{Step B} is feasible;

{\bf STEP 4:} ``Generate" a list of all permutations of the elements in $Y$;

{\bf STEP 5:} Find a permutation ($Y_p$) from the above list such that for a fixed set $X$, the

{\hspace{1.5cm} new set $Y_pX$ obtained by joining $X$ to $Y_p$, denoted by $Y_pX$ (with the parameters

{\hspace{1.5cm} $L_{Y_pX}=1+\sum_{x_i\in X}x_i+\sum_{y_i\in Y_p}y_i\leq L$ and $m_{Y_pX}= \lfloor\#Y_pX\cdot\frac{m}{n-1}\rfloor$), allows a
\hspace*{2.1cm} small optimal step $\sigma$, in the sense of Remark \ref{rsigma};

{\bf STEP 6:} If such a permutation, resulting in  a small value of $\sigma$, does not exist in Step 5,

\hspace*{1.5cm} then back to Step 3 and choose another set $Y$;

{\bf STEP 7:} Update the set $X\leftarrow Y_pX$, and repeat the steps 3 - 5 by adjoining new sets $Y_p$

\hspace*{1.5cm} until $\#Y_pX=n-1$;

{\bf STEP 8:} Return the set $D=Y_pX$.

\medskip
%\noindent
\begin{remark}\label{5.1}
%With the modified algorithm, we create the set $D$ "by parts", with respect to the \textbf{Step A}.
The parameters $L_X$ and $m_X$ are derived by computer simulations, where  $L_X$ essentially constrains the set $X$ and $m_X$ keeps the proportionality between the numbers $m, \#X$ and $\#D=n-1.$
\end{remark}
The main question which arises here is whether the performance of the mentioned algorithms above can be improved by using some of the new presented modes in the previous section. Unfortunately, the main reasons why $GFSGA$ can not be replaced by $GFSGA^*_{(1)}$ or $GFSGA^*_{(2)}$ are the following:
\begin{enumerate}
\item Apart from Remark \ref{Mrem1}, due to their  low dependency on the choice of tap positions, neither $GFSGA^*_{(1)}$ nor $GFSGA^*_{(2)}$  mode (through Proposition \ref{vs}) simply do not provide enough information that can be used to construct the tap positions with high resistance to GFSGA attacks in general.
% In other words, the GFSGA attacks with variable step of sampling have a tendency to use the tap positions, but not to highly depend on their placement.
It is clear that the presented algorithms above only give a (sub)optimal placement of tap positions due to impossibility to test exhaustively all permutations of $D$ and additionally to perform testing of all difference sets $D$ is infeasible as well. An optimal placement of tap positions providing the maximum resistance to GFSGA attacks leads us back to Open problem \ref{op1}.
\item One may notice that the main role of the constant step $\sigma$ used  in the design of the algorithm in \cite{SHT} is to reduce the repetition of bits in general, since $\sigma$ may take any value from $1$ to $L.$ This reduction of the repeated  bits is significantly larger when using the constant step than any variable step of sampling  in  $GFSGA^*_{(1)}$ or $GFSGA^*_{(2)}$, due to their specific definitions given by (\ref{qkk2}) and (\ref{si}).
\end{enumerate}
Notice that some criteria for tap selection  regarding the resistance to the inversion attacks were proposed in \cite{Gol96}. The difference between the first and last tap position should be near or equal to $L-1,$ which turns out to be an equivalent criterion of maximization of the sum $\sum^{n-1}_{i=1}d_i\leq L-1$, as  mentioned above. Generalized inversion attacks \cite{Gol2000} performed on filter generators, with the difference between the first and last tap position equal to $M$ $(=l_n-l_1),$ have the complexity approximately  $2^M$  \cite{Gol2000}. Hence, taking that $\sum^{n-1}_{i=1}d_i=L-1$, where $d_i=l_{i+1}-l_i$ (with tap positions $\mathcal{I}_0=\{l_1,\ldots,l_{n}\}$), one of the criteria which thwarts (generalized) inversion attacks  is easily satisfied.
%we satisfy one of useful criterions which provide a resistance to both GFSGA-like attacks and inversion attacks.

In addition,  one may also use  a $\lambda$th-order full positive difference set \cite{Gol96} for tap selection, that is, the set of tap positions $\mathcal{I}_0=\{l_1,\ldots,l_{n}\}$ with as small as possible parameter $\lambda=\max_{1\leq \sigma\leq M}|\mathcal{I}_0\cap (\mathcal{I}_0+\sigma)|$. If $\lambda=1$, then $\mathcal{I}_0$ is a standard full positive difference set. As illustrated in Table \ref{tabCC2}, to provide a high resistance to GFSGA-like attacks, the set of tap positions may be a $\lambda$th-order full positive difference set with higher values of $\lambda$. Note that in the case of inversion attacks, smaller  $\lambda$ is required. In other words, ($\lambda$th-order) full positive difference sets do not provide the same resistance to inversion attacks and GFSGA-like attacks, when the selection of tap positions is considered.

\subsection{Full positive difference sets versus algorithms in \cite{SHT}}\label{sec:cons}

In this section, we compare the performance of the  three GFSGA modes by applying these attacks to a cipher whose tap positions are chosen using the algorithms in  \cite{SHT} and in the case  the tap positions form suitably chosen full positive difference sets, respectively. We analyze the resistance to GFSGA attacks (using these two methods for tap selection) and conclude that  full positive difference sets do not give an optimal placement of tap positions, i.e., they do not provide a maximal resistance to GFSGA-like cryptanalysis.

It is not difficult to see that the set of rules valid for the algorithms in \cite{SHT} essentially  require that we choose a set of tap positions $\mathcal{I}_0$ so that the corresponding set of consecutive differences $D$, apart from  having different elements (possibly all), is also characterized by the property that these differences are coprime and in a specific order.
%One may consider this as a necessary condition when it come to taps selection.
The situation when  taps are not chosen optimally, implying a high divisibility of the elements in $D$,  is illustrated in Table \ref{T1}.  Denoting by $T_{Comp.}$, $T^*_{Comp._{(1)}}$ and $T^*_{Comp._{(2)}}$  the running time of the  $GFSGA,$ $GFSGA^*_{(1)}$ and $GFSGA^*_{(2)}$ mode, respectively, it is obvious that GFSGA is superior to other modes in most of the cases, as indicated in Table~\ref{T1}.
\begin{table*}[h]
\scriptsize
\centering
\caption{Complexity comparision of all three GFSGA modes for "bad" tap choices.}
\vskip 2.5mm
\begin{tabular}{ccclll}
  \hline  \noalign{\smallskip}
 $L$ & $(n,m)$ & $D$ & $T_{Comp.}$ & $T^*_{Comp._{(1)}}$ & $T^{*}_{Comp._{(2)}}$ \\ \noalign{\smallskip} \hline \noalign{\smallskip}
   80 & (9,2)  & \{12, 3, 6, 12, 6, 4, 24, 12\} & $2^{43.97}$ & $2^{67.97}$ & $2^{62.97}$ \\ \noalign{\smallskip} \hline \noalign{\smallskip}
  120 & (11,3) & \{5, 10, 15, 4, 5, 10, 5, 15, 20, 25\} & $2^{37.7}$ & $2^{63}$ & $2^{69.7}$\\ \noalign{\smallskip} \hline \noalign{\smallskip}
  160 & (15,6) & \{14, 7, 3, 14, 7, 7, 14, 7, 14, 28, 7, 14, 14, 7\} & $2^{32.97}$ & $2^{32.97}$ & $2^{50.97}$  \\ \noalign{\smallskip} \hline \noalign{\smallskip}
%  256 & (26,9) & \{27, 6, 3, 9, 12, 15, 3, 18, 6, 9, 3, 24, 18, 3, 12, 9, 6, 12, 3, 9, 12, 3, 6, 3, 12\} & $2^{53}$  & $2^{52}$ & $2^{59}$\\  \noalign{\smallskip}
%  \hline
\end{tabular}
\label{T1}
\end{table*}
In Table \ref{tabCC}, we compare the performance of the three modes, if the tap positions (sets of differences $D$) are chosen suboptimally according to rules and algorithms in \cite{SHT}.
\begin{table*}[h]%\label{tap1}
\scriptsize
\centering
\caption{Complexity comparison of  GFSGA modes - algorithmic selection of taps.}
\vskip 2.5mm
\begin{tabular}{ccclll}
  \hline  \noalign{\smallskip}
 $L$ & $(n,m)$ & $D$ & $T_{Comp.}$ & $T^*_{Comp._{(1)}}$ & $T^{*}_{Comp._{(2)}}$ \\ \noalign{\smallskip} \hline \noalign{\smallskip}
   80 & (7,2)  & \{5, 13, 7, 26, 11, 17\} & $2^{69.97}$ & $2^{63.97}$ & $2^{59.97}$ \\ \noalign{\smallskip} \hline \noalign{\smallskip}
  120 & (13,3) & \{5, 7, 3, 13, 6, 11, 5, 11, 7, 13, 21, 17\} & $2^{99.7}$ & $2^{104}$ & $2^{78.7}$\\ \noalign{\smallskip} \hline \noalign{\smallskip}
  160 & (17,6) & \{5,11,4,3,7,9,1,2,23,15,5, 13, 7, 26, 11, 17\} & $2^{86.97}$ & $2^{79.97}$ & $2^{41.97}$  \\ \noalign{\smallskip} \hline \noalign{\smallskip}
  200 & (21,7) & \{3, 7, 9, 13, 18, 7, 9, 1, 2, 9, 1, 2, 23, 15, 5, 13, 7, 26, 11, 17\} & $2^{108.9}$ & $2^{96.93}$ & $2^{68.93}$\\ \noalign{\smallskip} \hline \noalign{\smallskip}
 % 256 & (27,9) & \{5, 9, 13, 4, 7, 19, 3, 7, 9, 13, 18, 7, 9,
 %  1, 2, 9, 1, 2, 23, 15, 5, 13, 7, 26, 11, 17\} & $2^{135}$  & $2^{107}$ & $2^{109}$\\  \noalign{\smallskip}
 % \hline
\end{tabular}
\label{tabCC}
\end{table*}
Thus, if the tap positions are chosen according to the rules and algorithms in \cite{SHT},
%the constant step easily draws a lot of repeated bits, and therefore $GFSGA$ is considered as the most effective mode on such taps. On the other hand,
it turns out that $GFSGA^{*}_{(1)}$ and $GFSGA^{*}_{(2)}$ modes are more efficient than $GFSGA.$

In Table~\ref{tabCC2}, we compare the resistance of a nonlinear filter generator (specified by $L$, $n$ and $m$) to different GFSGA  modes regarding the design rationales behind  the choice of tap positions. Namely, for the same cipher (in terms of the parameters above), the attack complexities are evaluated  for tap positions that form (suitable) full positive differences sets and,   respectively, for the choices of    tap positions given in Table \ref{tabCC} (with a slight modification adopted for different parameters $L, n$ and $m$). In general, the algorithmic approach  gives  a higher resistance to GFSGA-like cryptanalysis.
\begin{table*}[h]
\scriptsize
\centering
\caption{Complexity comparision - full positive difference sets versus  algorithmic choice.}
\vskip 2.5mm
\begin{tabular}{cccclll}
  \hline  \noalign{\smallskip}
 $L$ & $(n,m)$ & Tap positions - Full positive difference sets & & $T_{Comp.}$ & $T^*_{Comp._{(1)}}$ & $T^{*}_{Comp._{(2)}}$ \\ \noalign{\smallskip} \hline \noalign{\smallskip}
   80 & (7,2)  & \{1, 3, 8, 14, 22, 23, 26\}  & & $2^{35.97}$ & $2^{37.97}$ & $2^{57.97}$ \\ \noalign{\smallskip} \hline \noalign{\smallskip}
  120 & (13,3) & \{1, 3, 6, 26, 38, 44, 60, 71, 86, 90, 99, 100, 107\}   & & $2^{86.72}$ & $2^{90.72}$ & $2^{95.72}$\\ \noalign{\smallskip} \hline \noalign{\smallskip}
  160 & (15,4) & \{1, 5, 21, 31, 58, 60, 63, 77, 101, 112, 124, 137, 145, 146, 152\} & & $2^{96.97}$ & $2^{105.97}$ & $2^{116.97}$  \\ \noalign{\smallskip} \hline \noalign{\smallskip}
  200 & (17,5) & \{1, 6, 8, 18, 53, 57, 68, 81, 82, 101, 123, 139, 160, 166, 169, 192, \
200\} & & $2^{113.93}$ & $2^{123.93}$ & $2^{132.93}$\\ \noalign{\smallskip} \hline \noalign{\smallskip}
 % 256 & (19,6) & \{1, 2, 7, 26, 33, 73, 101, 109, 121, 131, 154, 170, 188, 191, 205, \
%232, 234, 243, 247\} & & $2^{125}$  & $2^{133}$ & $2^{151}$\\  \noalign{\smallskip}
 % \hline
  \hline  \noalign{\smallskip}
 $L$ & $(n,m)$ & Set of consecutive differences $D$ -algorithmic choice & $\lambda$ & $T_{Comp.}$ & $T^*_{Comp._{(1)}}$ & $T^{*}_{Comp._{(2)}}$ \\ \noalign{\smallskip} \hline \noalign{\smallskip}
   80 & (7,2)  & \{5, 13, 7, 26, 11, 17\} & 1 & $2^{69.97}$ & $2^{63.97}$ & $2^{59.97}$ \\ \noalign{\smallskip} \hline \noalign{\smallskip}
  120 & (13,3) & \{5, 7, 3, 13, 6, 11, 5, 11, 7, 13, 21, 17\} & 3 & $2^{99.7}$ & $2^{104}$ & $2^{78.7}$ \\ \noalign{\smallskip} \hline \noalign{\smallskip}
  160 & (15,4) & \{5, 3, 7, 1, 9, 17, 15, 23, 5, 13, 7, 26, 11, 17\} & 3 & $2^{114.97}$ & $2^{124.97}$ & $2^{101.97}$  \\ \noalign{\smallskip} \hline \noalign{\smallskip}
  200 & (17,5) & \{7, 13, 10, 13, 7, 1, 9, 17, 15, 23, 5, 13, 7, 26, 11, 17\} & 3 & $2^{120.93}$ & $2^{120.93}$ & $2^{113.93}$\\ \noalign{\smallskip} \hline \noalign{\smallskip}
 % 256 & (19,6) & \{17, 21, 23, 13, 10, 13, 7, 1, 9, 17, 15, 23, 5, 13, 7, 26, 11, 17\} & 4 & $2^{138}$  & $2^{139}$ & $2^{126}$\\  \noalign{\smallskip}
 % \hline
\end{tabular}
\label{tabCC2}
\end{table*}
\begin{remark}\label{IA}
Table \ref{tabCC2} also indicates that  an algorithmic choice of tap positions may provide significantly better resistance against GFSGA-like attacks compared to full positive difference sets (for various parameters $n,m$ and $L$).
%However, it does not mean that one should avoid difference sets as tap positions (in certain situations) for the given LFSR and consider them as "bad taps". As mentioned earlier, ($\lambda$th-order) full positive difference sets as tap positions provide a high resistance to inversion attacks.
%Namely, to attain a high resistance to powerful inversion attacks, ($\lambda$th-order) full positive difference sets have been proposed as tap positions in \cite{Gol96}.
\end{remark}
}

\subsection{Further examples and comparisons}

In this section we provide a few examples which illustrate the sampling procedure and specification of repeated bits for the $GFSGA^*_{(1)}$ and $GFSGA^*_{(2)}$ modes. In both cases we consider the set of consecutive differences $D=\{5,13,7,26,11,17\}$ (most of the differences being prime numbers) which corresponds to the set of tap positions $\mathcal{I}_0=\{1,6,19,26,52,63,80\}.$ We first consider the $GFSGA^*_{(1)}$ mode.
\begin{example}\label{80} Let the set of tap positions be given by $\mathcal{I}_0=s^{t_1}=\{1, 6, 19, 26, 52, 63, 80\},$ where $L=80$ and $F:GF(2)^{7}\rightarrow GF(2)^{2}$ ($n=7,$ $m=2$). The set $\mathcal{I}_0$ is chosen according to the algorithms in \cite{SHT}  and it is most likely an optimal choice of tap positions for the given parameters $L,$ $n$ and $m.$ Recall that the variable sampling steps  $\sigma_i$ for $GFSGA^*_{(1)}$ are determined by the maximum function used in relation (\ref{qkk2}). In Table \ref{tabcssx}, using the relation (\ref{qkk2}) we identify the repeated state bits until the inequality $nc^*>L+R^*$ is satisfied for some $c^*.$
\begin{table}[h]
\scriptsize
\centering
\vskip 2.5mm
\begin{tabular}{cccc}
  \hline
 $i$ & Sets $\mathcal{I}^*_i$ (where $(k+1)\leftrightarrow s_k$) & $q_i$ & $\sigma_i$ \\ \hline
  1 & \{6\} & 1 & 5  \\ \hline
 2 & \{19,24\} & 2 & 13 \\ \hline
 3 & \{26, 31, 44\} & 3 & 7  \\ \hline
 4 & \{52, 57, 70, 77\} & 4 & 26  \\ \hline
 5 & \{63, 68, 81, 88, 114\} & 5 & 11  \\ \hline
 6 & \{80, 85, 98, 105, 131, 142\} & 6 & 17  \\ \hline
 7 & \{85, 103\} & 2 & 5 \\ \hline
 8 & \{114, 147\} & 2 & 11  \\ \hline
 9 &\{131, 164, 175\} & 3 & 17  \\ \hline
 10 &\{118, 136\} & 2 & 5 \\ \hline
 11 & \{125, 138\} & 2 & 2  \\ \hline
  12 &\{131, 136, 182\} & 3 & 11  \\ \hline
 13 & \{138, 143, 156\} & 3 & 7   \\ \hline
14 &  \{164, 169, 182, 189\} & 4 & 26 \\ \hline
 15 & \{175, 180, 193, 200, 226\} & 5 & 11  \\ \hline
 16 & \{192, 197, 210, 217, 243, 254\} & 6 & 17  \\\hline
 17 &  \{197, 215\} & 2 & 5  \\ \hline
  18 &  \{199, 217\} & 2 & 2  \\ \hline
    19 & \{210, 215, 261\} & 3 & 11 \\ \hline
     20 & \{217, 222, 235\} & 3 & 7 \\ \hline
     21 &  \{243, 248, 261, 268\} & 4 & 26  \\
  \hline
\end{tabular}
\caption{Repeated bits attained by sampling steps $\sigma_i$ defined by (\ref{qkk2}).}
\label{tabcssx}
\end{table}
The total number of repeated equations over all observed outputs is $R^*=\sum^{c^*-1}_{k=1}q_k=67,$ where the number of outputs is $c^*=22.$ Note that the first observed output $w^{t_1}$ has the preimage space of full size, and thus there are no repeated bits. Since we chose $s^{t_1}=\mathcal{I}_0$, the positions of repeated bits at the corresponding tap positions can be found and calculated as follows:
\begin{itemize}
\item The step $\sigma_1=5$ gives the maximal intersection between $s^{t_1}$ and $s^{t_2}=\{s^{t_1}+5\}=\{6, 11, 24, 31, 57, 68, 85\}$, i.e., we have $$\max_{1\leq\sigma_1 \leq 80}\#\{s^{t_1}\cap (s^{t_1}+\sigma_1)\}=\max_{1\leq\sigma_1 \leq 80}\#\{s^{t_1}\cap s^{t_2}\}=\{6\},$$ which yields $q_1=1.$ The size of the preimage space is $|S_{w^{t_2}}|=2^{n-m-q_1}=2^{4}.$
\item Assuming the knowledge of $x^{t_2}\in S_{w^{t_2}}$ and $x^{t_1}\in S_{w^{t_1}},$ we search for an optimal  shift $\sigma_2$  of $s^{t_2}$ so that  $q_2=\#\{\{s^{t_1}\cup s^{t_2}\}\cap \{s^{t_2}+\sigma_2\}\}$ is maximized. Note that at this point, $\{s^{t_1}+\sigma_1\}=s^{t_2}$ is fixed. This gives $\sigma_2=13$ and $q_2=2.$ The set of repeated  bits is $$\max_{1\leq\sigma_2 \leq 80}\#\{\{s^{t_1}\cup s^{t_2}\}\cap \{s^{t_2}+\sigma_2\}\}=\{19, 24\},$$ since $s^{t_3}=\{s^{t_2}+\sigma_2\}=\{19, 24, 37, 44, 70, 81, 98\}.$
%Clearly, the starting positions of these bits can be seen from states $s^{t_1}$ and $s^{t_2}$.}
The preimage space has the cardinality $|S_{w^{t_3}}|=2^{n-m-q_2}=2^3.$
\end{itemize}
In this way, we can completely determine the preimage spaces and the positions of the repeated bits. Since for $i\in\{5,6,15,16\}$ we have $q_i\geq n-m=5$, then $2^{n-m-q_i}=1$ (by convention). Once the other values $q_j$  have been computed, for $j\in \{1,2,\ldots,21\}\backslash \{5,6,15,16\},$ the attack complexity  can be estimated as
$$T^*_{Comp._{(1)}}=2^{n-m}\times 2^{n-m-q_1}\times\ldots \times2^{n-m-q_{21}} \times L^3\approx  2^{63.97}.$$
%\textbf{The part below is given just to make a comparision between $GFSGA^*_{(1)}$ and $GFSGA$, since the title of this section is given as "Further examples and comparisons". If we give examples, without any comparisions, then we need to change the title. \textcolor[rgb]{0.98,0.00,0.00}{Now to reply on the comment about the empty sets  $\mathcal{I}_i$.} It is not important to have the case that all sets $\mathcal{I}_i$ are non-empty. It is important to take the optimal step of the attack for which the minimal complexity is obtained, regarding what the sets  $\mathcal{I}_i$ are. Any other choice of the step $\sigma$ different than $1,13,37$ will mean the higher complexity!!!!}\\\\
In the case of  GFSGA, an optimal choice of the sampling distance  (cf. Remark \ref{optstep}) is any $\sigma\in\{1,13,37\}.$ Each of these sampling steps requires $c=16$ observed outputs and gives $R=24$ repeated equations, where the set of all repeated bits is given by $$\{r_1,r_2,\ldots,r_{15}\}=\{0,0,0,0,1,1,2,2,2,2,3,3,4,4,4\}.$$
The values $r_i=0,$ $i=1,2,3,4,$ mean that the corresponding sets $\mathcal{I}_i$ are empty. The attack complexity of $GFSGA$ is then estimated as $T_{Comp.}\approx 2^{69.97}.$
\end{example}
\begin{example}
Now, for the same function $F$ and the set of tap positions $\mathcal{I}_0=s^{t_1}$ (or the set of differences $D=\{5,13,7,26,11,17\}$), we illustrate the $GFSGA^*_{(2)}$ mode. In Table \ref{NTX}, we show all repeated bits for the sampling steps $\sigma_i$ of the $GFSGA^*_{(2)}$ mode, which are defined by relation (\ref{si}). Recall that the steps $\sigma_i$ in this case are defined so that every $n-1=6$ outputs are at distances $d_i\in D.$
\begin{table}[h]
\scriptsize
\centering
\caption{Repeated bits attained by sampling steps $\sigma_i$ defined by (\ref{si}).}
\vskip 2.5mm
\begin{tabular}{cccc}
  \hline
 $i$ & Sets $\mathcal{I}^*_i$ (where $(k+1)\leftrightarrow s_k$)  & $q_i$ & $\sigma_i$ \\ \hline
 1 & \{6\} & 1 & 5 \\ \hline %&  \rdelim\}{6}{0mm}[First set of $6$ steps]  \\ \hline
 2 & \{19,24\} & 2 & 13 \\ \hline
 3 & \{26, 31, 44\} & 3 & 7 \\ \hline
 4 & \{52, 57, 70, 77\} & 4 & 26 \\ \hline
 5 & \{63, 68, 81, 88, 114\} & 5 & 11 \\ \hline
 6 & \{80, 85, 98, 105, 131, 142\} & 6 & 17 \\ \hline
 7 & \{85, 103\} & 2 & 5 \\ \hline %&  \rdelim\}{6}{0mm}[Second set of $6$ steps] \\ \hline
 8 & \{98, 103\} & 2 & 13 \\ \hline
9 & \{105, 110, 123\} & 3 & 7 \\ \hline
10 & \{131, 136, 149, 156\} & 4 & 26 \\ \hline
 11 & \{142, 147, 160, 167, 193\} & 5 & 11 \\ \hline
 12 & \{159, 164, 177, 184, 210, 221\} & 6 & 17 \\ \hline
 13 & \{164, 182\} & 2 & 5 \\ \hline %&  \rdelim\}{6}{0mm}[Third set of $6$ steps] \\ \hline
 14 & \{177, 182\} & 2 & 13 \\ \hline
 15 & \{184, 189, 202\} & 3 & 7 \\ \hline
 16 & \{210, 215, 228, 235\} & 4 & 26 \\\hline
 17 &  \{221, 226, 239, 246, 272\} & 5 & 11 \\ \hline
  18 &  \{238, 243, 256, 263, 289, 300\} & 6 & 17 \\ \hline
   19 &  \{243, 261\} & 2 & 5 \\ \hline %&  \rdelim\}{3}{0mm}[Part of Fourth set of  $6$ steps] \\ \hline
    20 &  \{256, 261\} & 2 & 13 \\ \hline
      21 & \{263, 268, 281\} & 3 & 7 \\
  \hline
\end{tabular}
\label{NTX}
\end{table}
By formula (\ref{cvs}), the complexity of $GFSGA^{*}_{(2)}$ is estimated as $T^*_{Comp._{(2)}}\approx 2^{59.97}$, and thus this mode outperforms both $GFSGA$ and $GFSGA^*_{(1)}$. The total number of repeated equations in this case is given by $R^{*}=72$,  for  $c^{*}=22$ observed outputs. Notice that both modes $GFSGA^*_{(1)}$ and $GFSGA^*_{(2)}$ required in total $22$ outputs  to construct  an overdefined system of linear equations $(nc^*>L+R^*)$.
\end{example}
\section{Employing GFSGA in other settings}\label{sec:applic}
The main limitation of GFSGA-like attacks is their large complexity when applied to standard filtering generators that only output a single bit  each time the cipher is clocked. In addition, this generic method cannot be applied in a straightforward manner in the cryptanalysis of ciphers that use NFSRs.
In this section, we discuss the possibility of improving  the efficiency and/or applicability  of GFSGA with variable sampling step for the above mentioned scenarios.
%We focus on two popular designed primitives, namely single output nonlinear filter function $f(x_1,...,x_n)$, ($m=1$) or nonlinear feedback shift register (NLFSR) driver.
It will be  demonstrated that GFSGA with variable sampling step may be employed in combination with other cryptanalytic methods to handle these situations as well.

\subsection{GFSGA applied to  single-output nonlinear filter generators ($m=1$)}
The time complexity of GFSGA with variable sampling step  is given by (\ref{cvs}), i.e.,
\begin{eqnarray}\label{cvs2}
T^*_{Comp.}&\hskip -2.5mm =\hskip -2.5mm &2^{n-m}\times 2^{n-m-q_1}\times \ldots\times 2^{n-m-q_{c^*-1}}\times L^3,\nonumber
\end{eqnarray}
and clearly when $m=1$ the complexity  becomes quickly larger than the time complexity of exhaustive search (for some common choices of the design parameters $n$ and $L$).
Based on  annihilators in fewer variables of a nonlinear filtering function $f(x_1,\ldots,x_n)$, Jiao {\em et al.} proposed  another variant of FSGA in \cite{Jiao2013}.
The core idea of this attack is to reduce  the size of the preimage space via annihilators in fewer variables $g_1(x_{j_1},\ldots,x_{j_d})$ and $g_2(x_{j_1},\ldots,x_{j_d})$ such that $f(x_1,\ldots,x_n)g_1(x_{j_1},\ldots,x_{j_d})=0$ and $(f(x_1,\ldots,x_n)\oplus 1) g_2(x_{j_1},\ldots,x_{j_d})=0$, where $\{j_1,\ldots,j_d\}\subset \{1,\ldots,n\}$. It was shown that the time complexity of this attack is given by
\begin{eqnarray}\label{cvs3}
T^{\Delta}_{Comp.}&\hskip -2.5mm =\hskip -2.5mm & {||S_{g_1=0}||}^{c_1}\times {||S_{g_2=0}||}^{c_2}\times L^\omega,\nonumber
\end{eqnarray}
where ${||S_{g_i=0}||}$, for $i=1,2$, is the size of  preimage space of the annihilator $g_i$ (restricted to the variables $\{j_1,\ldots,j_d\} $), $c^*= \lceil\frac{L}{d}\rceil$ is the number of sampling steps, $c^*= c_1+c_2$ and  $\omega=\log_2 7\approx2.807$ is the exponent of Gaussian elimination.
In \cite{Jiao2013}, it was also shown that this variant of FSGA could be applied to single-output filter generators. For instance, letting $L=87$ and  using a nonlinear Boolean functions $f(x_1,\ldots,x_6)$ as in ``Example 2" in  \cite{Jiao2013},
 it was demonstrated that the time complexity of this attack is only about $2^{80}$ operations, whereas the time complexity of  FSGA is about $2^{87}$ operations in \cite{Jiao2013}.

In a similar manner, the same approach leads to a reduction of time complexity when GFSGA with variable sampling step is considered.
For instance, let the set of tap positions be given by $\mathcal{I}_0=s^{t_1}=\{1, 6, 19, 26, 52, 63\}$ corresponding to the   inputs  $\{x_1, x_2, x_3, x_4, x_5, x_6\}$, respectively. Using the filtering function $f:\FF_2^6 \rightarrow \FF_2$ of ``Example 2" in  \cite{Jiao2013}, one can deduce that  $||S_{g_1(x_2,x_4,x_6)=0}||=||S_{g_2(x_2,x_4,x_6)=0}||=5$. Actually, we can only
consider the tap positions $\{6, 26, 63\}$ with full positive difference set $\{20, 37\}$. Moreover, let us use the  variable sampling steps $\sigma_i=20$ and $\sigma_{i+1}=37$, alternately.
In such a case, the preimage space of annihilator can be further reduced to $||S^*_{g_1(x_2,x_4,x_6)=0}||=||S^*_{g_2(x_2,x_4,x_6)=0}||\approx 2.5$ by using  the repeated bits.
The time complexity of GFSGA with variable sampling steps is about
$5\times {2.5}^{42}\times87^{2.807}\approx 2^{76.32}<2^{80}$ operations. In particular, the number of variable sampling points is $43$ since at the first sampling point 3 linear relations are obtained and the remaining 42 sampling points give
$2\times 42=84$ linear relations, thus in total  $3+84=87=L$ linear equations are derived. It directly means that our GFSGA with variable sampling step outperforms the variant of FSGA in \cite{Jiao2013}.

Due to the small sized parameters $L$ and $n$ the above example does not  illustrate a full potential   of using  GFSGA in cryptanalysis of single-output filtering generators. Its purpose is rather to show that GFSGA and its variants can be efficiently combined with other cryptanalytic methods. The most promising approach seems to be an interaction of GFSGA with algebraic attacks using small degree annihilators of  restrictions of the filtering function $f$. Indeed, the use of repeated bits not only reduces the preimage space it essentially also fixes a subset of input variables and therefore these restrictions of $f$ may have annihilators of very low degree. This implies the existence of additional low degree equations in state bits which may be either used for checking the consistency of the linear system and after all (for sufficiently large number of fixed variables) these equations become linear. It is beyond the scope of this paper to investigate further  the performance of this combined method but we believe that this kind of attack may become efficient against single-output filter generators with standard choice of the parameters $L$ and $n$.

\subsection{Applying GFSGA to NFSR-based ciphers}

A current tendency in the  design of stream ciphers, motivated by efficient hardware implementation, is the use of NFSRs in combination with (rather simple) nonlinear filtering function. For instance, this idea was employed in the design of the  famous stream ciphers Trivium \cite{Can2006} and Grain family  \cite{Agren2011}.
Apparently,  none of the GFSGA variants  can be applied  for recovering the initial state of these ciphers  but rather  for deducing certain internal state  of the cipher.
In this scenario,  the complexity of GFSGA is directly related to the complexity of solving  an overdefined
system of low  degree equations rather than a system of linear equations. More precisely,
assuming that the length of NFSR  is $L$ bits, the algebraic degree of its update function  is $r$, and the filtering
 function  $F:GF(2)^n \rightarrow GF(2)^m$,  then the time complexity of  GFSGA  is given by
\begin{eqnarray}\label{cvs5-1}
T^{**}_{Comp.}&\hskip -2.5mm =\hskip -2.5mm &2^{n-m}\times 2^{n-m-q_1}\times \ldots\times 2^{n-m-q_{c^*-1}}\times D^\omega,
\end{eqnarray}
where $D=\sum_{i=0}^{e\times r}\binom{L}{i}$, $\omega=2.807$ is the coefficient of Gaussian elimination, $c^*$ is the number of sampling steps,  and $e$ is closely related to the parameters $n, m, L, c^*$ and specified tap positions. The complexity being much larger than for LFSR-based ciphers, due to the term $D^\omega$, makes GFSGA methods practically inefficient.

However, one may mount another mode of internal state recovery attack which employs the GFSGA sampling procedure, but without solving systems of equations at all.  More precisely, this new type of internal state recovery attack  also employs  the sampling of outputs within a certain sampling window  which then allows us to efficiently recover a certain portion of internal state bits from the reduced preimage spaces  corresponding  to the observed outputs.
To describe the attack in due detail, let us denote by $p$  the distance  between the last entry of NFSR (where NFSR is updated) and the  tap position  closest  to this registry cell. We assume that this distance satisfies the inequality $(p-1)\times n>L$, where $n$ is the number of inputs  (tap positions) of a filtering function $F:\FF_2^n \rightarrow \FF_m$. Note that this condition  implies that either $p$ or $n$ are relatively large. In such a case, let us choose the constant sampling steps $\sigma_i=1$, for $i=1,\ldots, p-1$, and assume the adversary can directly recover, say $R_p$ internal state bits (in total), at these $p-1$ sampling instances. The remaining $L-R_p$ internal state bits are still unknown and  the adversary can exhaustively guess these bits to recover the whole internal state. The process of identifying the correct internal state is as follows. For each possible internal state candidate, a portion of  $L$ keystream bits $Z^{t}=(z_1,\ldots,z_L)$ at time instance $t$ is determined using a given encryption algorithm. Then, comparing $L$   keystream bits $Z^{*t}=(z_1^*,\ldots,z_L^*)$ at time instance $t$ generated by the  cipher (with unknown secret internal state), we can distinguish the correct  and wrong internal states by directly checking if $Z^{t}=Z^{*t}$. In particular, if $Z^{t}=Z^{*t}$, then the guessed internal state would be the correct one, otherwise  another internal state candidate is considered. Consequently, the time complexity of this internal state recovery attack, assuming that remaining $L-R_p$ bits are guessed,  is given by
\begin{eqnarray}\label{cvs6-1}
T^{**}_{Comp.}&\hskip -2.5mm =\hskip -2.5mm &2^{n-m}\times 2^{n-m-q_1}\times \ldots\times 2^{n-m-q_{p-2}}\times 2^{L-R_p}.
\end{eqnarray}
The memory complexity of this attack is only $(p-1)\times n \times 2^{n-1}+L$ bits, which are used to save all the element of preimage spaces and $L$ keystream bits.
%\textbf{As I understand, we are observing exactly $p-1$ outputs, and thus we have that $q_i$s go to up to $q_{p-1}$?! In the last version it was up to $q_{c^*-1},$ where $c^*$ was not defined...}\\\\

The following example illustrates an application of this approach to an NFSR-based cipher that largely resembles the NFSR used in the Grain-128 cipher. In particular, the process of recovering $R_p$ internal state bits is described more thoroughly.
%\textcolor[rgb]{0.00,0.00,1.00}{, since we will use modified update function of the NFSR and set of tap positions which correspond to difference set.}
%The following example illustrates an application of GFSGA to a NLFSR-based cipher whose design shares the same structure and components with Grain-128 cipher. The only different point is
%the key length of this variant Grain-128 cipher uses an $L=256$ bits master key rather than an $L=128$ bits key used in standard Grain-128 cipher.
%{\bf I do not understand this example !!!  Why not consider Grain-128 since its design is so close to the considered one ??  We cannot apply this approach to Grain-128 because of cubic and quartic terms, right ? Is there any means for treating these terms as well ? What about LFSR there, do we attack just NFSR with this approach ot the whole Grain cipher ? The probability that these 3 terms in Grain-128 are  zero is quite large, so maybe we can propose probabilistic attack ! Also we need to give some more details about the complexity computation as I am sure that the reviewers will request some explanations !!}
%
%\textcolor[rgb]{1.00,0.00,0.00}{\textbf{
%Yes! Your comment is very good! I will try to explain it below clearly. }}
\begin{example}\label{ex:85}
Let $L=128$, $n=8, m=1$. The update function
of NFSR is defined below (a slightly modified variant of the NFSR used in Grain-128  \cite{Agren2011} without cubic and quartic terms):
\begin{eqnarray*}b_{t+128}&=&1 \oplus b_t \oplus b_{t+26} \oplus b_{t+56} \oplus b_{t+91} \oplus b_{t+96} \oplus b_{t+3}b_{t+67} \oplus b_{t+11}b_{t+13} \\
&& \oplus b_{t+17}b_{t+18} \oplus b_{t+27}b_{t+59} \oplus b_{t+40}b_{t+48} \oplus b_{t+61}b_{t+65} \oplus b_{t+68}b_{t+84}.
\end{eqnarray*}
%{\bf ( without cubic and quartic terms)} \\
The set of tap positions which we consider is given by $\mathcal{I}_0=s^{t_1}=\{l_1,\ldots,l_8\}=\{1, 7, 21, 26, 52, 67,$\\ $ 89, 105\},$ and it corresponds to a full positive difference set $\{6, 14, 5, 26, 15, 22, 16\}.$
The distance between the last tap position  and the NFSR update  position is $p=128-105=23$. This means
that if we consider an updated internal state bit (the first nonlinear bit) and constant sampling rate $\sigma_i=1$, this bit will appear at the tap position after 23 sampling instances.
% it will be shifted to the last tap position $l_8$ in $23$ sampling steps if $\sigma_i=1$.
At the same time, employing the fact that many of these bits appear at some tap positions (thus using the idea of GFSGA), the adversary  directly obtains many bits corresponding to  $128$-bit internal state as follows:
\begin{enumerate}[1.]
\item By relation (\ref{ii}) and sampling steps described in Section \ref{sec:varstep}, collecting all repeated bits over $p-1$ observed outputs (which are on consecutive distances $\sigma_i=1$), we determine all preimage spaces $S_{w^{t_i}}$ ($i=1,\ldots,p-1$) and their sizes.

\item
%Since in the previous computation we were taking (consecutively) arbitrary vectors from each preimage space $S_{w^{t_i}},$ so that we have their bits in disposal, then exactly at the time instance $t_{p-1}$ in registry state there are many bits which are repeated (accumulated) from previous states (which actually come from preimage spaces).
 Our approach implies that at the  sampling instance $i$  we recover (essentially guess)  $n-q_i$ internal state bits which must match to one of the $2^{n-q_i-1}$ preimages.
% total number of recovered bits denote by $R_p.$
  It is important to note that the bits which come from preimage spaces are the only candidates to be an internal state of the registry, since they are precisely determined by consecutive repetitions over $p-1$ observed outputs.
%\textcolor[rgb]{1.00,0.00,0.00}{It is important to note that bits which come from preimage spaces are the only candidates to be an internal state of the registry, since they are precisely determined by consecutive repetitions over $p-1$ observed outputs.} \textbf{This red sentence is the most important one in the whole approach, since it is the main reason why we only need to guess just $L-R_p$ remaining bits, i.e., all other $R_p$ bits are determined by repetitions over all observed outputs !! Even more precisely, $R_p$ is not just the number of repeated bits (over observed registry states), but it is the number of all different bits (equations) which came from preimage spaces on all consecutive observed registry states. That is the reason why we say "...the total number of recovered bits denote by $R_p$", and Wei (referring to $R_p$) is using the term "recovered internal bits" or "internal bits" only..}

\end{enumerate}
  Table \ref{tab:NTX2} specifies the number of recovered internal state bits, and the sizes of corresponding preimage spaces.
Denoting by $R_p$ the total number of recovered bits, we can see (from Table \ref{tab:NTX2})  that $R_p=8+8\times4 +7+6\times8+5+4+3\times6=122<128,$ where these bits are calculated using (\ref{ii}), for $\sigma_i=1,$ $(i=1,\cdots, 22).$

The adversary can further guess the remaining $L-R_p=128-122=6$ internal state bits, and thus the time complexity, using (\ref{cvs6-1}), of this attack   is about
$$T^{**}_{Comp.}=2^{7+7\times4+6+5\times8+4+3+2\times6} \times 2^{6}=2^{106}< 2^{128}.$$
The data complexity of this attack is only about $22+128=150$ keystream bits. The memory complexity is upper bounded by $22\times8\times 2^7+128 <2^{15}$ bits, which corresponds  to storing at most $2^7$ elements from preimage spaces and $128$  keystream bits. The success rate is close to one since there are $2^{7+7\times4+6+5\times8+4+3+2\times6+6}=2^{106}$ internal state candidates in total and therefore only a small portion of about $2^{106} \times 2^{-128}=2^{-22}<1$ wrong internal state candidates  can pass the test.
%Therefore, the succuss rate of this attack is near one.
\end{example}

%{\bf How do we get these numbers 24, 5, 32, .... ??}\\
Example \ref{ex:85} demonstrates that GFSGA-like attacks can be applied} to NFSR-based stream ciphers without employing any structural properties of the filtering function.
%However, the specification of GFSGA  on practical stream ciphers such as Grain family (or Trivium) fully employing the choice of tap positions,  nonlinear terms of NLFSR, and the properties of the nonlinear filtering  functions  remains an interesting research topic.
\begin{table}[h]
\scriptsize
\centering
\caption{Recovered bits obtained by sampling step $\sigma_i=1$}
\label{tab:NTX2}
\vskip 2.5mm
\begin{tabular}{cccc}
  \hline
 $i$ &Recovered bits of internal state &$q_i$& The size of  preimage space \\ \hline
 1 & 8 & 0 & $2^7$\\ \hline %&  \rdelim\}{6}{0mm}[First set of $6$ steps]  \\ \hline
 2 & 8 & 0 & $2^7$\\ \hline
 3 & 8 & 0 & $2^7$\\ \hline
 4 & 8 & 0 & $2^7$\\ \hline
 5 & 7 & 1 & $2^6$\\ \hline
 6 & 6 & 2 & $2^5$\\ \hline
 7 & 6 & 2 & $2^5$\\ \hline %&  \rdelim\}{6}{0mm}[Second set of $6$ steps] \\ \hline
 8 & 6 & 2 & $2^5$\\ \hline
9 & 6 & 2 & $2^5$\\ \hline
10 & 6 & 2 & $2^5$\\ \hline
 11 & 6 & 2 & $2^5$\\ \hline
 12 &6 & 2 & $2^5$\\ \hline
 13 & 6 & 2 & $2^5$\\ \hline
 14 & 5 & 3 & $2^4$\\ \hline
 15 & 4 & 4 & $2^3$\\ \hline
 16 & 3 & 5 & $2^2$\\ \hline
 17 & 3 & 5 & $2^2$\\ \hline
 18 &  3 & 5 & $2^2$\\ \hline
 19 & 3 & 5 & $2^2$\\ \hline
 20 & 3 & 5 & $2^2$\\ \hline
 21 & 3 & 5 & $2^2$\\ \hline
 %22 & 2 & 6 & $2$\\ \hline
 %23 & 2 & 6 & $1$\\ \hline
 %24 &  2 & 6 & $1$\\ \hline
 %25 &  2 & 6 & $1$\\ \hline
 %26 &  1 & 7 & $1$\\ \hline
 %27 &  1 & 7 & $1$\\ \hline
 %28 &  1 & 7 & $1$\\ \hline
 %29 &  1 & 7 & $1$\\ \hline
 %30 &  1 & 7 & $1$\\ \hline
 %31 &  1 & 7 & $1$\\ \hline
 %32 &  1 & 7 & $1$\\ \hline
\end{tabular}
%\label{NTX2}
\end{table}
%\textcolor[rgb]{1.00,0.00,0.00}{\begin{remark}\label{10}
%It is possible to further reduce the time complexity of this attack, if the adversary solves an overdefined system of low  degree equations (with remaining 12-variables) via the approach of equation (\ref{cvs5-1}) rather than an exhaustively search the remaining 12 bits internal state.
%\end{remark}}
The following example illustrates an application of GFSGA to a hybrid NFSR/LFSR-based cipher whose design is very similar to Grain-128 cipher. The major difference is the key length, since our variant assumes that the key length is  $L=256$ bits  rather than  $128$-bit key used in Grain-128 \cite{Agren2011}.
\begin{example}\label{ex:86}
Let $L=256$, $n=17, m=1$.
The internal state of our variant of Grain-128   consists of one 128-bit LFSR and one 128-bit NFSR, whose  state bits are denoted by $(s_0,\cdots, s_{127} )$ and  $(b_0,\cdots, b_{127})$, respectively. Their update functions  are  defined respectively as follows (see also   \cite{Agren2011}):
\begin{eqnarray}
s_{t+128} &=& s_t \oplus s_{t+7} \oplus s_{t+38} \oplus s_{t+70} \oplus s_{t+81} \oplus s_{t+96} \\
b_{t+128} &=& s_t \oplus b_t \oplus b_{t+26} \oplus b_{t+56} \oplus b_{t+91} \oplus b_{t+96} \oplus b_{t+3}b_{t+67} \oplus b_{t+11}b_{t+13} \nonumber \\
& & \oplus b_{t+17}b_{t+18} \oplus b_{t+27}b_{t+59} \oplus b_{t+40}b_{t+48} \oplus b_{t+61}b_{t+65} \oplus b_{t+68}b_{t+84} \nonumber
\end{eqnarray}
For this variant of Grain-128 cipher we consider the same set of tap positions that are used in the standard Grain-128 cipher, i.e.,  the tap position are $A= \{2, 12, 15, 36, 45, 64, 73, 89, 95\}$ for the  NFSR and $B= \{8, 13, 20, 42, 60, 79, 93, 95\}$ for the  LFSR. Note that the largest tap index in $A$ is $95$, and the NFSR is updated at position $128$, i.e., their distance is $p=128-95=33$. Similarly as in Example \ref{ex:85}, sampling at the constant rate $\sigma_i=1$, Table \ref{tab:NTX3} specifies the number of recovered (repeated) bits of internal state, and the size of preimage spaces.
\begin{table}[hbt]
\scriptsize
\centering
\caption{Repeated bits attained by sampling step $\sigma_i=1$}
\vskip 2.5mm
\begin{tabular}{cccc}
  \hline
 $i$ &Recovered bits of internal state &$q_i$& The size of preimage space \\ \hline
 1 & 17 & 0 & $2^{16}$\\ \hline %&  \rdelim\}{6}{0mm}[First set of $6$ steps]  \\ \hline
 2 & 16 & 1 & $2^{15}$\\ \hline
 3 & 15 & 2 & $2^{14}$\\ \hline
 4 & 15 & 2 & $2^{14}$\\ \hline
 5 & 14 & 3 & $2^{13}$\\ \hline
 6 & 13 & 4 & $2^{12}$\\ \hline
 7 & 12 & 5 & $2^{11}$\\ \hline
 8 & 12 & 5 & $2^{11}$\\ \hline
9 & 10 & 7 & $2^{9}$\\ \hline
10 & 9 & 8 & $2^{8}$\\ \hline
 11 & 9 & 8 & $2^{8}$\\ \hline
 12 &9 & 8 & $2^{8}$\\ \hline
 13 & 9 & 8 & $2^{8}$\\ \hline
 14 & 8 & 9 & $2^{7}$\\ \hline
 15 & 8 & 9 & $2^{7}$\\ \hline
 16 & 7 & 10 & $2^{6}$\\ \hline
 17 & 7 & 10 & $2^{6}$\\ \hline
 18 & 6 & 11 & $2^{5}$\\ \hline
 19 & 4 & 13 & $2^{3}$\\ \hline
 20 & 4 & 13 & $2^{3}$\\ \hline
 21 & 3 & 14 & $2^{2}$\\ \hline
 22 & 2 & 15 & $2$\\ \hline
 23 & 2 & 15 & $2$\\ \hline
 24 & 2 & 15 & $2$\\ \hline
 25 & 2 & 15 & $2$\\ \hline
 26 & 2 & 15 & $2$\\ \hline
 27 & 2 & 15 & $2$\\ \hline
 28 & 2 & 15 & $2$\\ \hline
 29 & 2 & 15 & $2$\\ \hline
 30 & 2 & 15 & $2$\\ \hline
 31 & 2 & 15 & $2$\\ \hline
% 32 & 2 & 15 & $2$\\ \hline
    & $\sum=227$  &    & $\prod=2^{196}$ \\ \hline
\end{tabular}
\label{tab:NTX3}
\end{table}

Thus, the adversary can directly obtain $17+227=244<256$ internal state bits. The  remaining $L-R_p=256-244=12$ internal state bits can then be guessed,
%{\bf (The use of GFSGA is not clear here !! Easier to guess exhaustevily !?)} \textcolor[rgb]{1.00,0.00,0.00}{\textbf{(Wei's comment: You are right! We do not need to mention the basic idea of GFSGA  here)}},
which  then leads to a recovery of the whole $256$-bit internal state. Therefore, the time complexity of this attack is about
$$T^{**}_{Comp.}=2^{16+196} \times 2^{12}= 2^{224}< 2^{256}.$$

\end{example}

The above example demonstrates that GFSGA-like cryptanalysis is also applicable to hybrid NFSR/LFSR-based ciphers.
In particular, it is shown that the tap positions have a very important impact on the security of NFSR/LFSR-based ciphers.

\begin{remark}\label{11}
In difference to the time-memory-data trade-off attacks or algebraic attacks, this attack has more favorable data  and memory complexity. For instance, in Example \ref{ex:86}, the data complexity of this attack is only  about $32+229=261\approx 2^{8}$  keystream bits. Namely, in the first step we use 32 sampling instances to determine all specified preimage spaces and their sizes under constant sampling rate $\sigma_i=1$, and
%In fact, there are $2^{197} \times 2^{27}= 2^{224}$  possible internal state candidates in total.
in the second step we need to use about $229$ fresh keystream bits to determine the correct state.
%In this case, a small portion of about $2^{224} \times 2^{-229}=2^{-5}<1$ wrong internal state candidate can pass the test \textcolor[rgb]{1.00,0.00,0.00}{(i.e., the succuss rate of this attack is near one). Therefore, we need about $2^{8}$ keystream bits in this attack.
Notice that the memory complexity of this attack is only about $32\times 17 \times 2^{16}+256\approx 2^{25}$ bits.  On the other hand, if the filtering function is $f: GF(2)^{17}\rightarrow GF(2)^m, m> 4$, then the time complexity of this attack is less than $2^{128}$ operations. It implies that this attack would outperform the time-memory-data trade-off attack for $m>4$.
\end{remark}

\subsection{Grain-128 tap selection}
We have already remarked that the tap selection for both SOBER-t32 and SFINKS was not optimal with respect to their resistance to GFSGA cryptanalysis, see also \cite{SHT}. We show that the same is true when Grain-128 is considered, thus there exist better selections that ensure greater resistance to GFSGA-like cryptanalysis.

We assume that either LFSR or NFSR, whose tap positions are given in Example~\ref{ex:86}, of Grain-128 are used as state registers in a filter generator and we apply different modes of GFSGA to these schemes. In the case when the LFSR of Grain-128 is employed in such a scenario then the complexities of the three different modes of GFSGA are given as,
\begin{table}[H]
%\footnotesize
\centering
\caption{Time complexity of different modes of GFSGA on LFSR of Grain-128}
\vskip 2.5mm
\begin{tabular}{|c|c|c|}
  \hline
    $T_{Comp.}$ & $T^*_{Comp._{(1)}}$ & $T^*_{Comp._{(2)}}$ \\ \hline
    $2^{108}$ & $2^{125}$& $2^{118}$ \\ \hline
\end{tabular}
%\caption{The scheme of all differences for $D=\{3,4,1,2\}$.}
\label{tab:Grain128LFSR}
\end{table}
Using our algorithm for finding a (sub)optimal placement of tap positions, instead of using the set $A=\{2, 12, 15, 36, 45, 64, 73, 89, 95\}$, we find another set of tap positions given as $\{1, 16, 27, 54, 71, 95, 108, 127\}$ which gives the following complexities,
$$T_{Comp.}=2^{129}, \;\;\; T^*_{Comp._{(1)}}=2^{132}, \;\;\; T^*_{Comp._{(2)}}=2^{123}.$$
%\begin{table}[H]
%%\footnotesize
%\centering
%\caption{GFSGA modes on LFSR of Grain-128 using taps $\{1, 16, 27, 54, 71, 95, 108, 127\}$}
%\vskip 2.5mm
%\begin{tabular}{|c|c|c|}
%  \hline
%  % after \\: \hline or \cline{col1-col2} \cline{col3-col4} ...
%    $T_{Comp.}$ & $T^*_{Comp._{(1)}}$ & $T^*_{Comp._{(2)}}$ \\ \hline
%    $2^{108}$ & $2^{125}$& $2^{118}$ \\ \hline
%
%\end{tabular}
%%\caption{The scheme of all differences for $D=\{3,4,1,2\}$.}
%\label{tab:Grain128LFSR}
%\end{table}
A similar improvement can also be achieved when the tap positions of NFSR in Grain-128 are considered. In this case the original placement of  tap positions (the set $B$ in Example~\ref{ex:86}) gives the following complexities,
\begin{table}[H]
%\footnotesize
\centering
\caption{Time complexity of different modes of GFSGA on NFSR of Grain-128}
\vskip 2.5mm
\begin{tabular}{|c|c|c|}
  \hline
    $T_{Comp.}$ & $T^*_{Comp._{(1)}}$ & $T^*_{Comp._{(2)}}$ \\ \hline
    $2^{114}$ & $2^{125}$& $2^{122}$ \\ \hline
\end{tabular}
%\caption{The scheme of all differences for $D=\{3,4,1,2\}$.}
\label{tab:Grain128NFSR}
\end{table}
On the other hand, our algorithm suggest somewhat better allocation of these taps given by $\{3, 10, 29, 42, 59, 67, 88, 103, 126\}$, which then induces the following complexities of the three GFSGA modes,

$$T_{Comp.}=2^{130}, \;\;\; T^*_{Comp._{(1)}}=2^{139}, \;\;\; T^*_{Comp._{(2)}}=2^{125}.$$

\section{Conclusions}\label{sec:conc}

In this article we have investigated the problem of selecting  tap positions of the driving LFSR used in filter generators. The importance of this problem seems to be greatly neglected by the designers since to the best of our knowledge only some heuristic design rationales (such as the concepts of full positive difference sets and co-primality of consecutive differences) can be found in the literature.
%These rationales (apart from being rather imprecise) do not ensure an optimal resistance to guess and determine cryptanalysis, in particular to the GFSGA modes discussed here.
The algorithmic approach of selecting taps (sub)optimally given their number and the length of LFSR appears to generate good solutions for this problems, though its further optimization and development may result in a better performance. Two additional modes of  GFSGA  have  been introduced and it turns out that these modes in many cases can outperform the standard GFSGA mode. The idea of  combining GFSGA technique and algebraic attacks appears   to be a promising unified cryptanalytic method against LFSR-based stream ciphers though a more thorough analysis and some practical applications are needed for confirming its full potential.

%{\bf \large{Acknowledgements:}}
\section*{Acknowledgment}
\addcontentsline{toc}{section}{Acknowledgment}  Samir Hod\v zi\' c  is supported in part by the Slovenian Research Agency (research program P3-0384 and Young Researchers Grant) and  Enes Pasalic is partly supported  by the Slovenian Research Agency (research program P3-0384 and research project J1-6720). Yongzhuang Wei
was supported in part by the Natural Science Foundation of China (61572148), in part by the Guangxi Natural
Science Found (2015GXNSFGA139007), in part by the project of Outstanding Young Teachers Training in
Higher Education Institutions of Guangxi.

%\newpage

{\Large {\bf Appendix}}\newline \newline
Since finding the preimage spaces $S_{w^{t_i}}$ of the observed outputs $w^{t_i}$ is the most important part, we give for clarity the description of a few initial steps:\\\\
%Since we are going to observe LFSR states, the bits of the sequence $(s_k)_{k\in \mathbb{N}}$ will be changed with the corresponding natural numbers, i.e. $s_k\rightarrow (k+1).$ (\textbf{Is it necessary to give more explanation on this notation?})
%%It means that we will work with natural number, and clearly they corresponds to some bits...
%Let the tap positions be given by the set $\mathcal{I}_0=\{i_1,i_2,\ldots,i_n\}.$
\textbf{Step 1:} Let $w^{t_1}$ denotes the first observed output so that the corresponding LFSR state at the tap positions is exactly the set $\mathcal{I}_0\hskip -1.5mm=\hskip -1.5mm \{l_1,l_2,\ldots,l_n\}$, so that $s^{t_1}\hskip -1.5mm=\hskip -1.5mm (s^{t_1}_{l_1}, \ldots,s^{t_1}_{l_n})\hskip -1.5mm \stackrel{(\ref{int})}{=}\hskip -1.5mm\{l_1,\ldots,l_n\}$, i.e., $s^{t_1}\hskip -1.5mm=\hskip -1.5mm \mathcal{I}_0$. Notice that the first observed output $w^{t_1}$ does not necessarily need to correspond to the set $\mathcal{I}_0,$ though (for simplicity) we assume this is the case.
%Clearly, it is not necessary to consider the output $w^{t_1}$ which corresponds to the set $\mathcal{I}_0,$ i.e., we can consider tap positions at any LFSR state.
A preimage space which corresponds to the first observed output $w^{t_1}$ always has the size $2^{n-m},$ i.e., $|S_{w^{t_1}}|=2^{n-m}.$ \newline \newline
%Now, taking the $n$-tuple $x^{t_1}=(x^{t_1}_1,x^{t_1}_2,\ldots,x^{t_1}_n)\in S_{w^{t_1}}$ we dispose of $n$ its bits.\\\\
\textbf{Step 2:}
%By the first taken $n$ bits $x^{t_1}=(x^{t_1}_1,x^{t_1}_2,\ldots,x^{t_1}_n)\in S_{w^{t_1}}$,
Taking the second output $w^{t_2}$ at distance $\sigma_1$ from  $w^{t_1}$ (thus $t_2=t_1+\sigma_1$),
%Note that after selecting the output $w^{t_2}$, the first step $\sigma_1$ is fixed. Using the set of LFSR state bits at the tap positions at time instance $t_1$, which is denoted by $s^{t_1}=\mathcal{I}_0$,
 we are able to identify  and calculate the number of repeated bits (equations) at the time instance $t_2$. Using the notation above, the set $\mathcal{I}^*_1$ of these bits is given by the intersection:
$$\mathcal{I}^*_1=s^{t_1}\cap s^{t_2}=s^{t_1}\cap \{s^{t_1}+\sigma_1\}=\mathcal{I}_0 \cap \{l_1+\sigma_1,\ldots,l_n+\sigma_1\},$$
where $s^{t_2}=\{s^{t_1}+\sigma_1\}\stackrel{\textnormal{def}}{=}(s^{t_1+\sigma_1}_{l_1+\sigma_1}, s^{t_2+\sigma_1}_{l_2+\sigma_1}, \ldots,s^{t_n+\sigma_1}_{l_n+\sigma_1})\stackrel{(\ref{int})}{=} \{l_1+\sigma_1,\ldots,l_n+\sigma\}$, i.e., $s^{t_2}=\{l_1+\sigma_1,\ldots,l_n+\sigma\}$ is the LFSR state at tap positions at time instance $t_2.$ Denoting the cardinality of $\mathcal{I}^*_1=s^{t_1}\cap s^{t_2}$ by $q_1,$ i.e., $q_1=\#\mathcal{I}^*_1,$ the cardinality of the preimage space  corresponding to the output $w^{t_2}$ is given as $|S_{w^{t_2}}|=2^{n-m-q_1}.$
%Hence, by finding the repeated bits at the time instance $t_2$ we actually have the set $S_{w^{t_2}}.$ Taking the $n$-tuple $x^{t_2}=(x^{t_2}_1,x^{t_2}_2,\ldots,x^{t_2}_n)\in S_{w^{t_2}},$ we dispose of its $n$ bits.\\\\
%\textbf{Step 3:} Since we have taken $x^{t_1}\in S_{w^{t_1}}$ and $x^{t_2}\in S_{w^{t_2}}$, we choose the third output $w^{t_3}$ on distance $\sigma_2$ to the output $w^{t_2}$, i.e., $t_3=t_2+\sigma_2=t_1+(\sigma_1+\sigma_2).$ The set $\mathcal{I}^*_2$ of repeated bits at the time instance $t_3$ is given as an intersection of the set $s^{t_3}$ and union of sets $s^{t_1}$ and $s^{t_2}$, i.e., we have:
%$$\mathcal{I}^*_2=\{s^{t_1}\cup s^{t_2}\}\cap s^{t_3}=\{s^{t_1}\cup s^{t_2}\}\cap \{s^{t_1}+(\sigma_1+\sigma_2)\}.$$
%The cardinality of the corresponding preimage space is given as $|S_{w^{t_3}}|=2^{n-m-q_2},$ where $q_2=\#\mathcal{I}^*_2.$ At this point steps $\sigma_1$ and $\sigma_2$ are fixed, since we have already chosen outputs $w^{t_1}$ and $w^{t_2}$ and determined their corresponding preimage spaces.\\\\

This process is then continued using the sampling distances $\sigma_2, \ldots, \sigma_{c^*}$ until the condition $nc^*-R^*>L$ is satisfied, where the total number of repeated equations over $c^*$ observed  outputs is $R^*=\sum^{c^*-1}_{k=1}q_k.$ Note that the number of repeated equations  corresponding to the first output is $0,$ since the corresponding LFSR state $s^{t_1}$ is the starting one. Therefore the sum goes to $c^*-1.$
%$$q_1=\#\{\mathcal{I}_0 \cap \{l_1+\sigma_1, l_2+\sigma_1, \ldots,l_n+\sigma_1\}\}=\#\{s^{t_1}\cap \{s^{t_1}+\sigma_1\}\},$$
%where $s^{t_1}=\mathcal{I}_0$ and $(s^{t_1}+\sigma_1)=\{l_1+\sigma_1, l_2+\sigma_1, \ldots,l_n+\sigma_1\}.$ In general, we denote $s^{t_k}=(s^{t_1}+\sum^{k}_{i=1}\sigma_i)=\{l_1+\sum^{k}_{i=1}\sigma_i,\ldots,l_n+\sum^{k}_{i=1}\sigma_i\},$ for $k\in \mathbb{N}.$ Once we have chosen $\sigma_1,$ it implies that the second output is $w^{t_2}$, so that $t_2=t_1+\sigma_1.$


\begin{thebibliography}{10}

\bibitem{Anders}
{\sc R. Anderson}.
\newblock Searching for the optimum correlation attack.
\newblock In {\em Fast Software Encryption, FSE 94}, vol. LNCS, pp. 137--143. Springe-Verlag, 1995.


\bibitem{Agren2011}
 {\sc M.~Agren, M.~Hell, T.~Johansson, and W.~Meier}.
 \newblock A new version of Grain-128 with optional authentication.
 \newblock {\em International Journal of Wireless and Mobile Computing}, vol.~5, no.~1, pp.~48--59, 2011.


\bibitem{BirySha}
{\sc A.~Biryukov and A.~Shamir}.
\newblock Cryptanalytic time/memory/data tradeoffs for stream ciphers.
\newblock In {\em Advances in Cryptology---ASIACRYPT~2000}, vol. LNCS~1976,
  pp.~1--13. Springer-Verlag, 2000.

\bibitem{AnBarkProb}
{\sc A.~Braeken and B.~Preneel}.
\newblock Probabilistic algebraic attacks.
\newblock In {\em IMA Conference on Cryptography and Coding}, vol. LNCS~3796,
  pp.~290--303. Springer-Verlag, 2005.

%\bibitem{CC02}
%{\sc C.~Carlet}.
%\newblock A larger class of cryptographic {B}oolean functions via a study of
%  the {M}aiorana-{M}c{F}arland constructions.
%\newblock In {\em Advances in Cryptology---CRYPTO~2002}, vol. LNCS~2442,
%  pp. 549--564. Springer-Verlag, 2002.

\bibitem{CourMO}
{\sc N.~Courtois}.
\newblock Algebraic attacks on combiner with memory and several outputs.
\newblock In {\em International Conference on Information Security and
  Cryptology -- ICISC 2004}, vol. LNCS~3506, pp. 3--20. Springer-Verlag,  2005.

\bibitem{CourM03}
{\sc N.~Courtois and W.~Meier}.
\newblock Algebraic attacks on stream ciphers with linear feedback.
\newblock In {\em Advances in Cryptology---EUROCRYPT~2003}, vol. LNCS~2656,
  pp.~346--359. Springer-Verlag, 2003.


\bibitem{Can2006}
{\sc C. D. ~Canni\`{e}re and B. ~Preneel}.
\newblock Trivium: A stream cipher construction inspired by block cipher design principles.
\newblock In \emph{Information Security}, vol. LNCS~4176,
pp.~171--186. Springer-Berlin Heidelberg, 2006.





\bibitem{Gol94}
{\sc J.~Dj. Goli\'{c}}.
\newblock Intrinsic statistical weakness of keystream generators.
\newblock In {\em Advances in Cryptology---ASIACRYPT~1994}, vol. LNCS~917,
  pp.~91--103. Springer-Verlag, 1995.

\bibitem{Gol96}
{\sc J.~Dj. Goli\'{c}}.
\newblock On the security of nonlinear filter generators.
\newblock In {\em Fast Software Encryption '96}, vol. LNCS~1039, pp.~  173--188. Springer-Verlag, 1996.

\bibitem{Gol2000}
{\sc J.~Dj. Goli\'{c}, Andrew Clark, and Ed~Dawson}.
\newblock Generalized inversion attack on nonlinear filter generators.
\newblock {\em IEEE Trans. Computers}, vol.~49, no.~10, pp.~1100--1109, 2000.

\bibitem{Hawkes}
{\sc P. Hawkes and G. Rose}.
\newblock Primitive specification and supporting documentation for SOBER-t16 submission to NESSIE.
\newblock {\em In Proceedings of the First Open NESSIE Workshop, KU-Leuven}, 2000.

\bibitem{HellmTMTO}
{\sc M.~Hellman}.
\newblock A cryptanalytic time-memory tradeoff.
\newblock {\em IEEE Trans. on Inform. Theory}, vol.~26, no.~4, pp.~401--406, 1980.

\bibitem{HongS05}
{\sc J.~Hong and P~Sarkar}.
\newblock New applications of time memory data tradeoffs.
\newblock In {\em Advances in Cryptology---ASIACRYPT~2005}, vol. LNCS~3788,   pp. 353--372. Springer-Verlag, 2005.



\bibitem{Jiao2013}
{\sc L.~Jiao, M.~Wang, Y.~Li, and M.~Liu}.
 \newblock On annihilators in fewer variables£º basic theory and applications.
\newblock {\em Chinese Journal of Electronics}, vol.~22, no.~3, pp.~489--494, 2013.



\bibitem{mass69}
{\sc J.~L. Massey}.
\newblock Shift-register synthesis and {BCH} decoding.
\newblock {\em IEEE Trans. on Inform. Theory}, vol.~15, no.~1, pp.~122--127, 1969.

\bibitem{Meier1}
{\sc W. Meier and O. Staffelbach}.
\newblock Fast correlation attacks on certain stream
ciphers.
\newblock  In \emph{J. of Cryptology}, vol. 1(3), pp. 159--176, 1989.



\bibitem{MPCEC04}
{\sc W.~Meier, E.~Pasalic, and C.~Carlet}.
\newblock Algebraic attacks and decomposition of {B}oolean functions.
\newblock In {\em Advances in Cryptology---EUROCRYPT~2004}, vol. LNCS~3027,  pp.~474--491. Springer-Verlag, 2004.

\bibitem{Handb}
{\sc A. J. Menezes, P. C. Van Oorschot,  S. A. Vanstone,  R. L. Rivest},
  \newblock Handbook of Applied Cryptography. 1997
\bibitem{Decim}
{\sc M. J. Mihaljevi\'c, M. P. C. Fossorier, and H. Imai}.
\newblock A general formulation of algebraic and fast correlation attacks
based on dedicated sample decimation.
\newblock In {\em Applied Algebra,
Algebraic Algorithms and
Error-Correcting Codes}, vol. LNCS~3857,  pp.~203--212, Springer-Verlag, 2006.

\bibitem{Norm2012}
{\sc M. J. Mihaljevi\'c, S. Gangopadhyay, G. Paul, and H. Imai}.
\newblock Internal state recovery of Grain-v1 employing normality order of the filter function
\newblock In {\em IET Information Security}, vol. 6, no. 2, pp.~55--64, 2006.

\bibitem{KaiNy92}
{\sc K.~Nyberg}.
\newblock On the construction of highly nonlinear permutations.
\newblock In {\em Advances in Cryptology---EUROCRYPT'92}, vol. LNCS~658,  pp.~92--98. Springer-Verlag, 1992.

%\bibitem{EnFSGA}
%{\sc E.~Pasalic}.
%\newblock On guess and determine cryptanalysis of {LFSR}-based stream ciphers.
%\newblock {\em IEEE Trans. on Inform. Theory}, vol.~55, no.~7, pp.~3398--3406, 2009.

%\bibitem{Hawkes}
%{\sc P. Hawkes and G. Rose}.
%\newblock Primitive specification and supporting documentation for SOBER-t16 submission to NESSIE.
%\newblock {\em In Proceedings of the First Open NESSIE Workshop, KU-Leuven}, 2000.

\bibitem{SHT}
{\sc E. Pasalic, S. Hod\v{z}i\'{c}, S. Bajri\'{c}, Y. Wei.}
\newblock Optimizing the placement of tap positions.
\newblock {\em Cryptography and Information Security in the Balkans---BalkanCryptSec 2014}, vol. LNCS 9024, pp. 15--30, Springer-Verlag 2015.


\bibitem{PAAEP09}
{\sc E.~Pasalic}.
\newblock Probabilistic versus deterministic algebraic cryptanalysis - a
  performance comparison.
\newblock {\em IEEE Trans. on Inform. Theory}, vol.~55, no.~11, pp.~2182--2191, 2009.

\bibitem{EnFSGA}
{\sc E.~Pasalic}.
\newblock On guess and determine cryptanalysis of {LFSR}-based stream ciphers.
\newblock {\em IEEE Trans. on Inform. Theory}, vol.~55, no.~7, pp.~3398--3406, 2009.

\bibitem{S85}
{\sc T. Siegenthaler}.
\newblock Decrypting a class of stream cipher using ciphertext only.
\newblock  \emph{IEEE Trans. Comp.}, vol. C-34, no.1, pp.81--85, January 1985.

\bibitem{EnesGFSGA}
{\sc Y.~Wei, E.~Pasalic and Y.~Hu}.
\newblock Guess and determinate attacks on filter generators--Revisited.
\newblock {\em IEEE Trans. on Inform. Theory}, vol.~58, no.~4, pp.~2530--2539, 2012.

\bibitem{Sfinx}
{\sc A. Braeken, J. Lano, N. Mentens, B. Preneel, and I. Verbauwhede.}
\newblock SFINKS: A synchronous stream cipher for restricted hardware environments.
\newblock {\em eSTREAM, ECRYPT Stream Cipher Project}, Report 2005/026 (2005)



\end{thebibliography}
\end{document}